\documentclass[%
 aip,
 amsmath,amssymb,
 reprint,%
]{revtex4-1}

\usepackage{graphicx}
\usepackage{dcolumn}
\usepackage{bm}

\usepackage[utf8]{inputenc}
\usepackage[T1]{fontenc}
\usepackage{mathptmx}
\usepackage{etoolbox}

\usepackage{graphicx}
\usepackage{epstopdf}
\usepackage{hyphenat}
\usepackage{amsmath}
\usepackage{setspace}
\usepackage{textcomp}
\usepackage{rotating}
\usepackage{braket}
\usepackage{amssymb}
\usepackage{xcolor}
\usepackage{bbold}
\usepackage{lmodern}
\usepackage[colorlinks=true,urlcolor=blue,citecolor=blue,linkcolor=red]{hyperref}

\makeatletter
\def\@email#1#2{%
 \endgroup
 \patchcmd{\titleblock@produce}
  {\frontmatter@RRAPformat}
  {\frontmatter@RRAPformat{\produce@RRAP{*#1\href{mailto:#2}{#2}}}\frontmatter@RRAPformat}
  {}{}
}%
\makeatother
\begin{document}

\preprint{AIP/123-QED}

\title{A model of quantum spacetime}
\author{Tommaso Favalli}
\email{favalli@lens.unifi.it}
\affiliation{QSTAR, INO-CNR and LENS, Largo Enrico Fermi 2, Firenze, Italy}%
 \affiliation{Universit\'a degli Studi di Napoli Federico II, Via Cinthia 21, Napoli, Italy}
  
\author{Augusto Smerzi}%
\affiliation{QSTAR, INO-CNR and LENS, Largo Enrico Fermi 2, Firenze, Italy}%

\date{\today}

\begin{abstract}
We consider a global quantum system (the \lq\lq Universe\rq\rq) satisfying a double constraint, both on total energy and total momentum. Generalizing the Page and Wootters quantum clock formalism, we provide a model of 3+1 dimensional, non-relativistic, quantum spacetime emerging from entanglement among different subsystems in a globally ``timeless" and  ``positionless" Universe. 
\end{abstract}

\maketitle

	\section{Introduction}
\label{Introduction}
The idea that time may emerge from quantum entanglement was first proposed in 1983 by Don Page and William Wootters \cite{pagewootters,wootters,pagearrow} to solve the, so called, “problem of time”. This arises in the context of canonical quantization of gravity where, according to the Wheeler-DeWitt equation, the whole universe should be in an eigenfunction of the total Hamiltonian \cite{dewitt,isham}. Page and Wootters (PaW) theory splits the total Hilbert space into two entangled subsystems, $C$ and $S$, where $C$ is the clock subspace of an appropriately chosen clock observable. Looking at the relative state \cite{everett} of the subsystem $S$, PaW show that the dynamical Schrödinger evolution can be recovered with respect to the clock values. This approach has recently attracted a large interest and has stimulated several generalizations (see for example Refs. \onlinecite{lloydmaccone,esp2,vedral,vedraltemperature,macconeoptempoarrivo,interacting,simile,simile2,leonmaccone,review,review2,nostro,nostro2,timedilation,scalarparticles,dirac,foti,brukner,wigner,indefinite,asimmetry,pinto1,pinto2}), including an experimental illustration \cite{esp1}. 
The PaW framework can be read as an internalization of the time reference frame, with the clock being an 
appropriately chosen physical system and time is considered as \lq\lq what is shown on a quantum clock\rq\rq.
We study here the possibility to extend this protocol in order to internalize, together, 
the temporal and the spatial reference frames. In this approach also space becomes an emerging property of entangled subsystems and the concept of position is recovered relative to \lq\lq what is shown on a quantum rod\rq\rq. 

Quantum reference frames for the spatial degree of freedom have been extensively studied in quantum information and quantum foundations (see for example Refs. \onlinecite{burnett,QRF1,QRF2,QRF3,QRF4,QRF5,QRF6,QRF7,QRF8,QRF9,QRF10,QRF11,QRF12,QRF13,QRF14,QRF15,QRF16,QRF17,QRF18,QRF19,QRF20,QRF21}). In the quantum gravity literature, it has been suggested that quantum reference frames are needed to formulate a quantum theory of gravity \cite{dewitt,afundamental,QG1,QG2}. In Ref. \onlinecite{change1} (see also Refs. \onlinecite{change2,change3,change4,change5,change6,change7,change8,change9}) has also been introduced a formalism for describing transformations to change the description between different quantum reference frames in various contexts. Just as the PaW mechanism has been extensively studied in order to describe the temporal degree of freedom, all these works have dealt only with internalization of the space reference frame. 
Only recently, in Ref. \onlinecite{giacomini}, has been introduced a fully relational formulation of a $1+1$ dimensional spacetime for the case of a system of $N$ relativistic quantum particles in a weak gravitational field.

In this work we first focus on space and we divide the total Hilbert space in two entangled subsystems, $R$ and $S$, where $R$ is the quatum rod that acts as a spatial reference frame for $S$. A generalization of the PaW mechanism for the spatial degree of freedom has already been addressed in Refs. \onlinecite{hoehn1,hoehn2,hoehn3} (see also Refs. \onlinecite{change2,change3}). Here we give our own version adopting and generalizing the approach outlined in Ref. \onlinecite{pegg} (see also Ref. \onlinecite{peggbar}). We consider indeed discrete spectra for the momentum operators and we take the spatial degree of freedom described by POVMs. This choice allow us to recover continuous values for the spatial degrees of freedom even if the momenta have discrete spectra (the generalization to the case of a continuous spectrum is also disussed). We therefore assume the Universe satisfying a constraint on total momentum $\hat{P}_{tot} \ket{\Psi} = 0$. Even if the global position is completely undetermined, a well-defined relative position emerges from the entanglement between the two subsystems $R$ and $S$. 
Finally we introduce an additional subsystem $C$ acting as a clock and we consider the Universe satisfying a double constraint: both on total momentum and total energy, that is $\hat{P}_{tot} \ket{\Psi} = 0$ and $ \hat{H}_{tot}\ket{\Psi} = 0$. We show that this framework can be implemented consistently and we thus provide a model of non-relativistic quantum spacetime emerging from entanglement. 
In order to facilitate the reading and simplify the notation, in Sections II, III and VI we consider a single spatial degree of freedom for the subsystems $R$ and $S$. In Section V we generalize the results to the case of $3+1$ dimensional spacetime and we discuss some examples.

	\section{Emergent Relative Position from Entanglement}
\label{space}
\subsection{General Framework}
We divide the total Hilbert space (the \lq\lq Universe\rq\rq) in two subsystems $R$ and $S$ where
$R$ acts as quantum reference frame for $S$. We consider the two subsystems non-interacting and entangled with global Hamiltonian
\begin{equation}
	\hat{H} = \hat{H}_R + \hat{H}_S
\end{equation}
where $\hat{H}_R$ and $\hat{H}_S$ act on $R$ and $S$ respectively. We consider the momenta of $R$ and $S$ having a discrete, bounded, non-degenerate spectra and introduce the spatial degrees of freedom as POVMs. The case of momenta with continuous, unbounded spectrum will be discussed in Section III.E. 

We begin by first considering the $R$ subspace. In introducing the POVMs of space we follow a generalization of the framework outlined in Ref. \onlinecite{pegg} (see also Refs. \onlinecite{nostro,nostro2}), namely we assume the momentum operator $\hat{P}_R$ of $R$ with point-like spectrum, equally-spaced eigenvalues and non-degenerate eigenstates. It can be illustrated by taking  $d_R$ momentum states $\ket{p_k}$ and $p_k$ momentum levels with $k=0,1,2,...,d_R -1$ such that (we set $\hslash=1$):

\begin{equation}\label{pk}
	p_k = p^{(R)}_0 + \frac{2\pi}{L_R} k .
\end{equation}
In this space we can define the states 

\begin{equation}\label{statixj}
	\ket{x_j}_R  = \frac{1}{\sqrt{d_R}}\sum_{k=0}^{d_R -1}e^{- i p_k x_j}\ket{p_k}_R
\end{equation}
and the values 

\begin{equation}\label{xdiscreti}
	x_j = x_0 + j \frac{L_R}{D_R} 
\end{equation} 
with $j=0,1,2,...,z=D_R -1$ and with the constraint $z+1=D_R \ge d_R$. If we take $D_R=d_R$ the states (\ref{statixj}) are orthogonal (with the $D_R=d_R$ values of $x_j$ uniformly distributed over $L_R$) and the spatial degree of freedom is described by the Hermitian operator $\hat{X}_R = \sum_{j}x_j\ket{x_j}\bra{x_j}$ complement of $\hat{P}_R$. If, instead, we take $D_R > d_R$ the number of the states $\ket{x_j}_R$ is greater than the number of momenta states in $R$ (with the $D_R$ values of $x_j$ again uniformly distributed over $L_R$). These states still satisfy the key property $\ket{x_j}_R = e^{- i\hat{P}_R(x_j - x_0) }\ket{x_0}_R$ and it can furthermore be used for writing the resolution of the identity:
\begin{equation}\label{pomidentity}
	\mathbb{1}_{R} = \frac{d_R}{D_R} \sum_{j=0}^{D_R -1} \ket{x_j}\bra{x_j}.
\end{equation}
Thanks to (\ref{pomidentity}) the spatial degree of freedom is represented by a POVM, being $d_R D^{-1}_{R} \ket{x_j}\bra{x_j}$ the $D_R$ non-orthogonal elements. In order to obtain a continuous representation of the coordinate $x$ (maintaining a discrete momentum spectrum), we can now consider the limit $z \longrightarrow \infty$, defining 
\begin{equation}\label{xstateinf}
	\ket{x}_R = \sum_{k=0}^{d_R -1 } e^{- i p_k x}\ket{p_k}_R
\end{equation}
where $x$ can now take any real value from $x_0$ to $x_0 + L_R$. In this limiting case the resolution of the identity (\ref{pomidentity}) becomes
\begin{equation}\label{newresolution}
	\mathbb{1}_{R} = \frac{1}{L_R} \int_{x_0}^{x_0+L_R} dx \ket{x} \bra{x} .
\end{equation}
The states $\ket{x_j}$ and $\ket{x}$ are not orthogonal, but we will see in the following that this will not constitute a problem in our derivation of spacetime. 

As mentioned, also the subspace $S$ can be equipped with the POVMs of space considering $\hat{P}_S$ with discrete, bounded spectrum and applying the same formalsim adopted in the subspace $R$. So we assume that also in $S$ all the momentum eigenvalues can be written as multiples of a minimum step, that is $p_k = p^{(S)}_0 + \frac{2\pi}{L_S} k$ with $k=0,1,2,...,d_S -1$. We thus define the states $\ket{y_l}_S  = \frac{1}{\sqrt{d_S}}\sum_{k=0}^{d_S -1}e^{-i p_k y_l}\ket{p_k}_S$ and the $z+1=D_S$ values $y_l = y_0 + l \frac{L_S}{D_S}$ or, in the limiting case ($z \longrightarrow \infty$) in which $y$ take any real value from $y_0$ to $y_0 + L_S$, we consider the states $\ket{y}_S = \sum_{k=0}^{d_S -1 } e^{- i p_k y}\ket{p_k}_S$. Also in this case, when taking $D_S=d_S$ we can define the operator $\hat{Y}_S = \sum_{l}y_l\ket{y_l}\bra{y_l}$ (complement of $\hat{P}_S$) which is an Hermitian operator.


\subsection{Emergent Relative Distance}
In order to obtain the emergence of space from entanglement we consider now the following constraint on the total momentum:
\begin{equation}\label{constmomentum}
	\hat{P}\ket{\Psi} = ( \hat{P}_R + \hat{P}_S)\ket{\Psi} = 0
\end{equation}
where $\hat{P}_R $ and $\hat{P}_S$ act on $R$ and $S$ respectively. Assuming $d_R \gg d_S$ the global state $\ket{\Psi}$ satisfying (\ref{constmomentum}) can be writen as
\begin{equation}\label{miserveperlafine}
	\ket{\Psi} = \sum_{k=0}^{d_S -1} c_k \ket{p=-p_k}_R\otimes\ket{p_k}_S .
\end{equation}
We can now expand $\ket{\Psi}$ in the $\left\{\ket{x_j}\right\}$ basis on $R$ through (\ref{pomidentity}), thus obtaining
\begin{equation}\label{stato1}
	\begin{split}
		\ket{\Psi} & =  \frac{d_R}{D_R} \sum_{j=0}^{D_R-1} \ket{x_j} \braket{x_j|\Psi} = \\& = \frac{\sqrt{d_R}}{D_R} \sum_{j=0}^{D_R-1} \ket{x_j}_R\otimes\sum_{k=0}^{d_S -1} c_k e^{- i p_k x_j}\ket{p_k}_S = \\&= \frac{\sqrt{d_R}}{D_R} \sum_{j=0}^{D_R-1} \ket{x_j}_R\otimes\ket{\phi(x_j)}_S 
	\end{split}
\end{equation}
where in last step we have defined $\ket{\phi(x_j)}_S = \sum_{k=0}^{d_S -1} c_k e^{- i p_k x_j}\ket{p_k}_S$. This state can be obtained from the global state $\ket{\Psi}$ through the \textit{relative state} definition \cite{everett} of the subsystem $S$ with respect to $R$ \cite{nostro}:
\begin{equation}\label{defstatorelativo}
	\ket{\phi(x_j)}_S = \frac{\braket{x_j|\Psi}}{1/\sqrt{d_R}}.
\end{equation}
Now using the fact that $\ket{x_j }_R = e^{- i\hat{P}_R (x_j -x_0)}\ket{x_0}_R$ and equations (\ref{constmomentum}) and (\ref{defstatorelativo}), we obtain
\begin{equation}\label{trasldiscreta}
	\begin{split}
		\ket{\phi(x_j)}_S & = \sqrt{d_R}\braket{x_j|\Psi} =\\&= \sqrt{d_R} \bra{x_0}e^{i\hat{P}_R (x_j -x_0)}\ket{\Psi} =  \\&
		=        \sqrt{d_R} \bra{x_0}e^{i (\hat{P} - \hat{P}_S) (x_j -x_0)}\ket{\Psi}  =\\&      = e^{- i \hat{P}_S (x_j -x_0)}\ket{\phi(x_0)}_S
	\end{split}
\end{equation}
that is, we have that the operator $\hat{P}_S$ is the generator of spatial traslations in the coordinate $x_j$. In this framework it is evident that the translation moves the system with respect to the coordinate of the reference frame $R$ and therefore \lq\lq external\rq\rq to $S$. Furthermore we want to discuss briefly the consequences of using states $\ket{x_j}_R$ in $R$ that are not orthogonal. Clearly this fact introduces a possible conceptual warning because considering $\ket{x_j}_R$ that are not orthogonal implies that these are partially indistinguishable with a single measurement, the probability of indistinguishability being proportional to $\left| \braket{x_i|x_j}\right|^2$. Nevertheless, as mentioned previously, this does not constitute a problem in our framework. Indeed, even if the $\ket{x_j}_R$ are partially indistinguishable, the state of the system $S$ conditioned on a given $x_j$ does not dependes on $x_i \ne x_j$ (as we can clearly see in equations (\ref{stato1}) and (\ref{trasldiscreta})) and so interference phenomena are not present even if the coordinates in $R$ are not orthogonal.

These results can easily extended in the limiting case $z \longrightarrow \infty$. Indeed in this case the global state satisfying the constraint (\ref{constmomentum}) can be written:
\begin{equation}\label{statoglobalecontinuo}
	\begin{split}
		\ket{\Psi} & =  \frac{1}{L_R} \int_{x_0}^{x_0 + L_R} d x   \ket{x} \braket{x|\Psi} = \\&= \frac{1}{L_R} \int_{x_0}^{x_0 + L_R} d x   \ket{x}_R \otimes \sum_{k=0}^{d_S-1} c_k e^{- i p_k x} \ket{p_k}_S =\\&= \frac{1}{L_R} \int_{x_0}^{x_0 + L_R} d x   \ket{x}_R \otimes \ket{\phi(x)}_S 
	\end{split}
\end{equation}
and, for the relative state $\ket{\phi(x)}_S = \braket{x|\Psi}$, can be easily obtained
\begin{equation}\label{traslcontinua}
	\begin{split}
		\hat{P}_S\ket{\phi(x)}_S &= \bra{x}\hat{P}_S\ket{\Psi} = \bra{x}(\hat{P} - \hat{P}_R)\ket{\Psi}=\\&  = - \bra{x}\hat{P}_R\ket{\Psi} =  
		 - \sum_{k=0}^{d_R -1} \bra{p_k}p_k e^{i p_k x} \ket{\Psi} =\\&= i\frac{\partial}{\partial x} \braket{x|\Psi} =  i\frac{\partial}{\partial x} \ket{\phi(x)}_S
	\end{split}
\end{equation}
that is the same of equation (\ref{trasldiscreta}), showing that the momentum $\hat{P}_S$ is the generator of translations in the coordinates $x$, but written through the differential expression $\hat{P}_S\ket{\phi(x)}_S = i\frac{\partial}{\partial x} \ket{\phi(x)}_S$.

In this framework the absolute position of $R+S$ is totally indeterminate. However, considering discrete values for the coordinates in $R$ and $S$, we can look for the conditional probability of obtaining $y_l$ on $S$ conditioned of having $x_j$ on $R$, where $y_l=y_0 + l\frac{L_S}{D_S}$, $x_j=x_0 + j\frac{L_R}{D_R}$. We have (see Appendix A for the proof)
\begin{multline}\label{conditionalprobabilitydiscreta}
	P(y_l \: on \: S \:|\: x_j \: on \: R) =\\=\frac{d_S}{D_S} \left|\braket{y_l|\phi(x_j)} \right|^2 = \frac{1}{D_S} \left| \sum_{k=0}^{d_S-1} c_k e^{i p_k(y_l-x_j)} \right|^2
\end{multline} 
that is a well-defined probability distribution, where $\sum_{l=0}^{D_S -1} P(y_l \: on \: S \:|\: x_j \: on \: R) = 1$. Working instead in the limit $z \longrightarrow \infty$, the probability for a value of $y$ in the small range between $y$ and $y + dy$ is given by $P(y \: on \: S \:|\: x \: on \: R)dy$, where the probability density $P(y \: on \: S \:|\: x \: on \: R)$ is (the proof is given in Appendix B):
\begin{multline}\label{conditionalprobability}
		P(y \: on \: S \:|\: x \: on \: R) =\\= \frac{1}{L_S} \left| \braket{y|\phi(x)} \right|^2  =\frac{1}{L_S} \left| \sum_{k=0}^{d_S-1} c_k e^{ i p_k(y-x)} \right|^2
\end{multline}
that is again a well-defined probability density distribution depending on the distance between $S$ and $R$. Indeed, also in this case, if we consider the integral over all possible values of $y$ we obtain 	
\begin{multline}
	\int_{y_0}^{y_0 + L_S} dy P(y \: on \: S \:|\: x \: on \: R) =\\=	\frac{1}{L_S} \int_{y_0}^{y_0 + L_S} dy \sum_{k}\sum_{n} c_k c^{*}_n e^{i(p_k - p_n)(y-x)} = 1
\end{multline}
where we have used (see Appendix C):
\begin{equation}\label{delta} 
	\int_{y_0}^{y_0 + L_S} dy e^{iy(p_k - p_n)} = L_S\delta_{p_k,p_n} .
\end{equation}
Equations (\ref{conditionalprobabilitydiscreta}) and (\ref{conditionalprobability}) display an essential feature of the complementarity between positions and momenta. Indeed, if the system $S$ is in an eigenstate of the momentum, in the right-hand side of (\ref{conditionalprobabilitydiscreta}) and (\ref{conditionalprobability}) there is only one term of modulus unity and we have $P(y_l \: on \: S \:|\: x_j \: on \: R) = 1/D_S$ and $P(y \: on \: S \:|\: x \: on \: R) = 1/L_S$. So, in this case, the probability $P(y_l \: on \: S \:|\: x_j \: on \: R)$ and the probability density $P(y \: on \: S \:|\: x \: on \: R)$ are constant across the whole interval $\left[y_0,y_0+L_S \right]$ indicating that, when the momentum of the system $S$ can be determined exactly, thus the position with respect to the reference frame $R$ is completely random. 

We have so considered a \lq\lq positionless\rq\rq Universe, satisfying the constraint (\ref{constmomentum}) on the total momentum (where the absolute position is totally indeterminate) and we found the well-defined conditional probability $P(y_l \: on \: S \:|\: x_j \: on \: R)$ (for the case of discrete coordinates) and the probability density $P(y \: on \: S \:|\: x \: on \: R)$ (for the limiting case $z \longrightarrow \infty$) where appears the relative distance between the two entangled subsystems.

\subsection{On the Position-Momentum Uncertainty Relation}
In this paragraph we want to focus on the quantity $\delta x$, namely the minimum interval in the $x$ values of $R$ over which the state of the system $\ket{\phi (x)}_S$ varies significantly. We will show that, if the momentum spread in the expansion (\ref{miserveperlafine}) is $\Delta p$, than $\delta x \ge \hslash /\Delta p$ (re-introducing $\hslash \ne 1$). This means that will be impossible to distinguish states of the system $S$ conditioned to $x$ values on $R$ which are close to each other less than $\simeq \hslash /\Delta p$, in accordance with position-momentum uncertainty relation. 

For the sake of simplicity we consider here $D_R=D_S=d_R=d_S = d$ and $L_R = L_S = L$. We assume also discrete values for the space, so we have $\ket{x_j}_R = \frac{1}{\sqrt{d}} \sum_{k}e^{-ip_kx_j} \ket{p_k}_R$ and $\ket{y_l}_S = \frac{1}{\sqrt{d}} \sum_{k}e^{- ip_k y_l} \ket{p_k}_S$ with $x_j=x_0 + j\frac{L}{d}$ and $y_l=y_0 + l\frac{L}{d}$. As already mentioned, in this particular case we can define the operators $\hat{X}_R = \sum_{j}x_j\ket{x_j}\bra{x_j}$ and $\hat{Y}_S = \sum_{l}y_l\ket{y_l}\bra{y_l}$ (complement of $\hat{P}_R$ and $\hat{P}_S$) which are Hermitian operators.

The crucial point for our argument here is to understand that in this framework, since the reference frame $R$ and the system $S$ are entangled in the global state $\ket{\Psi}$, the $\Delta p$ related to the spread of the coefficients in the expansion (\ref{miserveperlafine}), that is $	\ket{\Psi} = \sum_{k=0}^{d_S -1} c_k \ket{p=-p_k}_R\otimes\ket{p_k}_S$, does not refer exclusively to $R$, but to $R$ and $S$ together. For this reason what we will find is that a limited spread in the expansion in the momentum eigenbasis will reduce the distinguishability of states of $S$ conditioned on proximal values of $R$. So, starting from equation  
\begin{equation}\label{equazioneevoluzione2}
	\ket{\phi(x_j)}_S = \frac{\braket{x_j|\Psi}}{1/\sqrt{d}} = \sum_{k=0}^{d -1} c_k e^{- i \hslash^{-1} p_k x_j}\ket{p_k}_S,
\end{equation}
we can calculate in the space $S$:
\begin{equation}\label{ultima}
	\begin{split}
		\braket{\phi (x_i) | \phi (x_j)} 
		= \sum_{k=0}^{d-1} \left| c_k \right|^2 e^{- i\hslash^{-1} p_k (x_j - x_i)}. 
	\end{split}
\end{equation} 
Equations (\ref{ultima}) indicates that, if $\left| c_k \right|^2$ has a spread $\simeq \Delta p$, than the scalar product $f(x_j - x_i) = \braket{\phi (x_i) | \phi (x_j)}$ will have a spread of the order $\simeq \hslash/ \Delta p$. This means that the state $\ket{\phi(x_j)}_S$ of the subsystem $S$ varies significantly for intervals 
\begin{equation}
	\delta x  \geq \hslash/ \Delta p
\end{equation}
where $\Delta p$ is indeed the uncertainty in momentum related to the spread of the coefficients $c_k$. 
This means that it is impossible to distinguish states of the system $S$ conditioned to $x$ values on $R$ which are closer than $\simeq \hslash /\Delta p$ to each other, in accordance with the position-momentum uncertainty relation. 

Expression (\ref{ultima}) for $f(x_j - x_i)$ holds also in the case of non-orthogonal space states where the spatial degrees of freedom are described by POVMs. So we notice here that the function $f(x_j - x_i)$, and consequently the scale on which the system $S$ varies significantly, is not related to the overlap of the states in $R$. Namely the fact of using coordinates states that are not orthogonal does not have an effect on the derivation of the function $f(x_j - x_i)$. As already mentioned, this is because, calculating the conditioned state of $S$ to a certain value $x_j$ on $R$ through (\ref{equazioneevoluzione2}), we find no contributions from different positions $x_i \ne x_j$, and so interference phenomena are not present even if the position states are not orthogonal. Rather $f(x_j - x_i)$ is related to the spread of the coefficients appearing in the global state $\ket{\Psi}$ and this fact shows us a condition for a good functioning of our framework: a large spread in the coefficients within the global state in the expansion (\ref{miserveperlafine}) is needed in order to distinguish states of $S$ projected to closer values of $R$ \cite{asimmetry}. 


	\section{Spacetime from Entanglement}
\label{spacetime}	
\subsection{Introducing the $C$ Subspace}
In order to introduce the temporal degree of freedom we have to consider an additional Hilbert space and assign it to time. We so assume that the Universe is divided into three subsystems, that is we work in $\mathcal{H} = \mathcal{H}_C\otimes\mathcal{H}_R\otimes\mathcal{H}_S$, where $\mathcal{H}_C$ is the time Hilbert space. The three subsystems are non-interacting but still entangled, so we have 
\begin{equation}
\hat{H} = \hat{H}_C + \hat{H}_R + \hat{H}_S
\end{equation}
where $\hat{H}_C$, $\hat{H}_R$ and $\hat{H}_S$ act on $C$, $R$ and $S$ respectively. To be as general as possible, we consider the clock Hamiltonian $\hat{H}_C$ with bounded, discrete spectrum with unequally-spaced energy levels and we introduce also the time observable described by a POVM \cite{nostro,nostro2}. 

We start assuming $\hat{H}_C$ with point-like spectrum, non-degenerate eigenstates and having rational energy differences. 
The framewowrk can be illustrated by taking $p+1 = d_{C}$ energy states $\ket{E_i}$ and $E_i$ energy levels with $i=0,1,2,...,d_C -1$ such that $\frac{E_i -E_0 }{E_1 - E_0} = \frac{A_i}{B_i}$,	
where $A_i$ and $B_i$ are integers with no common factors. Doing this we obtain (we set again $\hslash=1$):
\begin{equation}\label{ei}
E_i = E_0 + r_i \frac{2\pi}{T}
\end{equation}
where $T=\frac{2\pi r_1}{E_1}$, $r_i = r_1\frac{A_i}{B_i}$ for $i>1$ (with $r_0=0$) and $r_1$ equal to the lowest common multiple of the values of $B_i$. In this space we define the states 

\begin{equation}\label{timestates}
\ket{t_m}_C  = \frac{1}{\sqrt{d_C}}\sum_{i=0}^{d_C -1 }e^{-i E_i  t_m}\ket{E_i}_C
\end{equation}
where 
\begin{equation}
	t_m = t_0 + m \frac{T}{D_C}
\end{equation}
with $m=0,1,2,...,s=D_C-1$ and $s+1 \ge r_p$ (we notice here that also in this case, by assuming the energy spectrum with equally-spaced eigenvalues and taking $D_C=d_C$, the time states (\ref{timestates}) are orthogonal and the temporal degree of freedom is described by the Hermitian operator $\hat{T}_C = \sum_{m}t_m \ket{t_m}\bra{t_m}$). The number of $\ket{t_m}_C$ states is therefore greater than the number of energy states in $\mathcal{H}_{C}$ and the $D_C$ values of $t_m$ are uniformly distributed over $T$. These states satisfy the key property $\ket{t_m}_C = e^{- i\hat{H}_C(t_m - t_0) }\ket{t_0}_C$ and furthermore can be used for writing the resolution of the identity in the $C$ subspace:
\begin{equation}\label{pomidentity2}
\mathbb{1}_{C} = \frac{d_C}{D_C} \sum_{m=0}^{D_C -1} \ket{t_{m}}\bra{t_{m}}.
\end{equation}
Thanks to property (\ref{pomidentity2}) also time is represented by a POVM, being $d_C D^{-1}_{C} \ket{t_m}\bra{t_m}$ the $D_C$ non-orthogonal elements. As we did for space, in order to obtain a continuous flow of time in the PaW framework, we can now consider the limit $s \longrightarrow \infty$, defining 

\begin{equation}\label{alphastateinf}
\ket{t}_C = \sum_{i=0}^{d_C-1} e^{- i E_i t}\ket{E_i}_C
\end{equation}
where $t$ can now take any real value from $t_0$ to $t_0 + T$. In this limiting case the resolution of the identity (\ref{pomidentity2}) becomes
\begin{equation}\label{newresolution2}
\mathbb{1}_{C} = \frac{1}{T} \int_{t_0}^{t_0+T} d t \ket{t} \bra{t} .
\end{equation}
As in the case of space, the states $\ket{t_m}_C$ and $\ket{t}_C$ are not orthogonal, but thanks to the properties (\ref{pomidentity2}) and (\ref{newresolution2}) they can be used as time states for introducing time through the PaW mechanism. This framework allows us to use a generic Hamiltonian as a clock Hamiltonian, with the only constraint of considering rational ratios of energy levels. However, we emphasize that this limitation can be opportunely relaxed. Indeed, in the case of non-rational ratios of energy levels, the resolutions of the identity (\ref{pomidentity2}) and (\ref{newresolution2}) are no longer exact but, since any real number can be approximated with arbitrary precision by a ratio between two rational numbers, the residual terms in the resolutions of the identity and consequent small corrections can be arbitrarily reduced. In this way we can consider that the mechanism works, at least approximately, for every generic Hamiltonian with no restrictions \cite{nostro,nostro2}.

\subsection{Emergent $1+1$ Dimensional Spacetime}
We want now to obtain a model of spacetime emerging from entanglement. So we consider the global state $\ket{\Psi} \in \mathcal{H}_C\otimes\mathcal{H}_R\otimes\mathcal{H}_S$ simultaneously satisfying

\begin{equation}\label{1}
\hat{H}\ket{\Psi} = (\hat{H}_C + \hat{H}_R + \hat{H}_S )\ket{\Psi}=0
\end{equation}
and
\begin{equation}\label{2}
\hat{P}\ket{\Psi} = (\hat{P}_R + \hat{P}_S )\ket{\Psi}=0
\end{equation}
where we have assumed $\hat{P}_C = 0$. We notice that the mechanism works also with $\hat{P}_C \ne 0$ but, in this case, there could be limitations in the allowed momenta to ensure that (\ref{1}) and (\ref{2}) are simultaneously satisfied (we discuss the case of $\hat{P}_C \ne 0$ in Appendix D). The framework with $\hat{P}_C = 0$ can be implemented for example by assuming the subspace $C$ describing the internal degree of freedom (i.e. the spin) of the system $R$. If we want to look at the explicit form of the state $\ket{\Psi}$ we need to know the relation between the momenta and the energy of $R$ and $S$. Nevertheless, assuming again $d_C, d_R \gg d_S$, we can write in general:
\begin{equation}\label{statoglobalespaziotempo}
\ket{\Psi} = \sum_{k=0}^{d_S -1} c_k \ket{E=-\epsilon_k}_C\otimes\ket{p=-p_k}_R\otimes\ket{p_k}_S
\end{equation} 
where 
\begin{equation}
\epsilon_k= E^{(R)}(-p_k) + E^{(S)}(p_k)
\end{equation}
is the energy function related to the momenta of $R$ and $S$. For simplicity we consider here that the energy function depends only on the momenta and not on the coordinates. Clearly the model works also considering the presence of external potentials in $R$ and $S$ (equations (\ref{evoluzioneRS}), (\ref{evoluzioneRSc}) and (\ref{m}) that we will find in the following would be indeed still valid) but, in this case, the state $\ket{\Psi}$ can not be written in the simple form (\ref{statoglobalespaziotempo}) and we can not explicitly calculate the conditional probabilities (\ref{probfinalediscreta}) and (\ref{probfinale}). For this reason we prefer to simplify the model, as we believe this choice helps to capture the essence of the mechanism. In Section III.E we give a generalization to the case where potentials are present in $R$ and $S$. So, considering for example $R$ and $S$ as free particles (with mass $M$ and $m$ respectively), we have $\hat{H}_R = \frac{\hat{P}^{2}_R}{2M}$ and $\hat{H}_S = \frac{\hat{P}^{2}_S}{2m}$, which implies $\epsilon_k = \frac{p^{2}_k}{2M} + \frac{p^{2}_k}{2m}$.

Starting from the state $\ket{\Psi}$ satisfying (\ref{1}) and (\ref{2}), we can now expand it on the basis $\left\{\ket{t_m}_C\right\}$ in $C$ thanks to (\ref{pomidentity2}), thus obtaining
\begin{equation}\label{serveperGLM}
\begin{split}
\ket{\Psi} &= \frac{d_C}{D_C} \sum_{m=0}^{D_C -1} \ket{t_m} \braket{t_m|\Psi} =\\&=  \frac{\sqrt{d_C}}{D_C} \sum_{m=0}^{D_C -1} \ket{t_m}_C \otimes \ket{\phi(t_m)}_{R,S}
\end{split}
\end{equation}
where $\ket{\phi(t_m)}_{R,S} = \sqrt{d_C} \braket{t_m|\Psi}$ is state of the composite system $R+S$ at time $t_m$, namely it is the \textit{relative state} \cite{everett} of $R+S$ conditioned on having the value $t_m$ on $C$. For such a state, through (\ref{1}) and the relative state definition, it is easy to find the time evolution with respect to the clock $C$:
\begin{equation}\label{evoluzioneRS}
\begin{split}
\ket{\phi(t_m)}_{R,S} &=\sqrt{d_C}\braket{t_m|\Psi} =\\&= \sqrt{d_C} \bra{t_0} e^{i\hat{H}_C(t_m - t_0)}\ket{\Psi}= \\&
= \sqrt{d_C} \bra{t_0} e^{-i (\hat{H}_R + \hat{H}_S - \hat{H})(t_m - t_0)}\ket{\Psi}=\\&= e^{-i (\hat{H}_R + \hat{H}_S)(t_m - t_0)}\ket{\phi(t_0)}_{R,S}
\end{split}
\end{equation} 
where $\ket{\phi(t_0)}_{R,S}= \sqrt{d_C} \braket{t_0|\Psi}$ is the state of $R+S$ conditioned on $t_0$ that is the value of the clock taken as initial time. Equation (\ref{evoluzioneRS}) shows, as expected, the simultaneous evolution of $R$ and $S$ over time. Having indeed considered a quantum spatial reference frame, it was reasonable to expect that it also evolves in time together with the subsystem $S$. We can consider then the limiting case $s \longrightarrow \infty$ where $t$ takes all the real values between $t_0$ and $t_0+T$. The global state can now be written 
\begin{equation}
\begin{split}
\ket{\Psi} = \frac{1}{T} \int_{t_0}^{t_0 +T} dt \ket{t} \braket{t|\Psi} = \frac{1}{T} \int_{t_0}^{t_0 +T} dt \ket{t}_C \otimes \ket{\phi(t)}_{R,S}
\end{split}
\end{equation}
and defining the relative state of $R+S$ as $\ket{\phi(t)}_{R,S} = \braket{t|\Psi}$ we obtain \cite{nostro}:
\begin{equation}\label{evoluzioneRSc}
i \frac{\partial}{\partial t}\ket{\phi(t)}_{R,S} = \left(\hat{H}_R + \hat{H}_S\right)\ket{\phi(t)}_{R,S}
\end{equation}
that is the Schrödinger evolution for the state of $R+S$ with respect to the clock time $t$, written in the usual differential form.

We can therefore expand the state $\ket{\Psi}$ in the coordinates $\left\{\ket{x_j}_R\right\}$ in $R$, thus obtaining:
\begin{equation}
\begin{split}
\ket{\Psi} &= \frac{d_R}{D_R} \sum_{j=0}^{D_R -1} \ket{x_j} \braket{x_j|\Psi} =\\&=  \frac{\sqrt{d_R}}{D_R} \sum_{j=0}^{D_R -1} \ket{x_j}_R \otimes \ket{\varphi(x_j)}_{C,S}
\end{split}
\end{equation}
where $\ket{\varphi(x_j)}_{C,S}=\sqrt{d_R}\braket{x_j|\Psi}$ is the relative state of $C+S$ conditioned to the value $x_j$ on the reference frame $R$. All the results found in the previous Section apply to the state $\ket{\varphi(x_j)}_{C,S}$. Indeed we have also in this case
\begin{equation}\label{m}
\ket{\varphi(x_j)}_{C,S} = e^{-i \hat{P}_S(x_j - x_0)} \ket{\varphi(x_0)}_{C,S}
\end{equation}
where the momentum of the clock $C$ does not appear since we have chosen $\hat{P}_C = 0$. Also here we consider the limit $z \longrightarrow \infty$, where again $x$ can take all the real values between $x_0$ and $x_0 + L_R$. In this case the global state can be written
\begin{multline}\label{mm}
\ket{\Psi} = \frac{1}{L_R} \int_{x_0}^{x_0 +L_R} dx \ket{x} \braket{x|\Psi} =\\= \frac{1}{L_R} \int_{x_0}^{x_0 +L_R} dx \ket{x}_R \otimes \ket{\varphi(x)}_{C,S}
\end{multline}
and defining the relative state of $C+S$ as $\ket{\varphi(x)}_{C,S} = \braket{x|\Psi}$ we obtain $\hat{P}_S \ket{\varphi(x)}_{C,S} = i \frac{\partial}{\partial x} \ket{\varphi(x)}_{C,S}$. Through this latter equation and (\ref{m}) we can see again that $\hat{P}_S$ is the generator of translations in the coordinate values $x$ for the relative state $\ket{\varphi(x)}_{C,S}$.

Finally we can expand the state $\ket{\Psi}$ simultaneously on the coordinates $\left\{\ket{x_j}_R\right\}$ in $R$ and on the time basis $\left\{\ket{t_m}_C\right\}$ in $C$. We have for the global state:
\begin{equation}\label{45}
\begin{split}
\ket{\Psi} &= \left(\frac{d_C}{D_C} \sum_{m=0}^{D_C -1} \ket{t_m}\bra{t_m} \otimes  \frac{d_R}{D_R} \sum_{j=0}^{D_R -1} \ket{x_j}\bra{x_j} \right)\ket{\Psi}=\\&
= \frac{\sqrt{d_C}}{D_C} \frac{\sqrt{d_R}}{D_R} \sum_{m=0}^{D_C -1}\sum_{j=0}^{D_R -1}\ket{t_m}_C\otimes\ket{x_j}_R\otimes\ket{\psi(t_m,x_j)}_S
\end{split}
\end{equation}
where $\ket{\psi(t_m,x_j)}_S =\sqrt{d_C} \sqrt{d_R}(\bra{t_m}\otimes\bra{x_j})\ket{\Psi}$ is the relative state of the system $S$ at time $t_m$ conditioned on the value $x_j$ for the reference frame $R$. With the state $\ket{\psi(t_m,x_j)}_S$ we have not yet defined the position of the system $S$. This state indeed gives us the value of the time that enters as a parameter thanks to the entanglement with the subspace $C$ and indicates the position of the reference frame $R$. What we can now search is the conditional probability of having a certain position $y_l$ in $S$ at time $t_m$ and knowing that the reference frame is in $x_j$, that is (see Appendix E):
\begin{multline}\label{probfinalediscreta}
P(y_l \: on\: S\:|\:x_j\:on\:R, \: t_m \: on \:C) =\\= \frac{d_S}{D_S} |\braket{y_l|\psi(t_m,x_j)}|^2 = \frac{1}{D_S} \left| \sum_{k=0}^{d_S -1} c_k e^{-i\epsilon_k t_m}e^{ip_k(y_l-x_j)} \right|^2
\end{multline}
where, we recall, $\epsilon_k$ is the energy function related to the momenta $p_k$ of $R$ and $S$ and where it is easy to verify that $\sum_{l=0}^{D_S -1} 	P(y_l \: on\: S\:|\:x_j\:on\:R, \: t_m \: on \:C) = 1$ given each $x_j$ and $t_m$. Clearly we can extend these results also to the limiting cases $z,s \longrightarrow \infty$. Indeed we can write the global state $\ket{\Psi}$ as
\begin{equation}\label{47}
\begin{split}
\ket{\Psi} &= \left( \frac{1}{T}\int_{t_0}^{t_0 + T} dt \ket{t}\bra{t} \otimes \frac{1}{L_R}\int_{x_0}^{x_0 + L_R} dx \ket{x}\bra{x} \right) \ket{\Psi} = \\&
= \frac{1}{T} \frac{1}{L_R} \int_{t_0}^{t_0 + T} dt \int_{x_0}^{x_0 + L_R} dx \ket{t}_C \otimes \ket{x}_R \otimes \ket{\psi(t,x)}_S
\end{split}
\end{equation}
where again $\ket{\psi(t,x)}_S = (\bra{t}\otimes\bra{x})\ket{\Psi}$ is the relative state of the system $S$ at time $t$ conditioned on the value $x$ for the reference frame $R$. The conditional probability density of having a certain position $y$ in $S$ at time $t$ and knowing that the reference $R$ is in $x$ is (see Appendix F):
\begin{multline}\label{probfinale}
P(y \: on\: S\:|\:x\:on\:R,\: t \: on \:C) =\\= \frac{1}{L_S}\left| \braket{y|\psi(t,x)} \right|^2= \frac{1}{L_S}\left| \sum_{k=0}^{d_S -1} c_k e^{-i\epsilon_k t}e^{ip_k(y-x)} \right|^2.
\end{multline}
We notice that also the probability density (\ref{probfinale}) is well-defined for each time (indeed it is easy to verify that $	\int_{y_0}^{y_0 + L_S} dy P(y \: on\: S\:|\:x\:on\:R,\: t \: on \:C) =1$ for all $x$ and $t$) and it depends on time $t$ as well as on the distance $y-x$ between $S$ and $R$.

So, through entanglement, we have found for the subsystem $S$ a conditional probability density that give us informations about the evolution of $S$ in time and space, where for time we consider the clock time and for space we consider the relative distance between $S$ and the quantum reference frame $R$. All these results are obtained within a globally static and \lq\lq positionless\rq\rq Universe.

To conclude this paragraph we notice that a good spatial reference frame is a reference that moves only slightly in time. If a good spatial reference frame is considered, one can look at the evolution of $S$ by itself. We show this point with an example: assuming $R$ and $S$ as free particles (with mass $M$ and $m$ respectively), we could take $M \gg m$ thus obtaining $\frac{\hat{P}^{2}_R}{2M} \ll \frac{\hat{P}^{2}_S}{2m}$. 
Starting from the state (\ref{47}) we can consider the relative state $\ket{\psi(t,x)}_S = (\bra{t}\otimes\bra{x})\ket{\Psi}$ and investigate its evolution. If the mass $M$ is sufficiently large, we have:
\begin{equation}\label{evS2}
i \frac{\partial}{\partial t}\ket{\psi(t,x)}_{S} \simeq \hat{H}_S \ket{\psi(t,x)}_{S} .
\end{equation}
Equation (\ref{evS2}) shows that if $M$ is sufficiently large, the evolution of $S$ alone can be recovered with respect to time $t$ and with respect to a spatial reference frame that does not evolve (or that move negligibly in time). Furthermore, equation (\ref{evS2}) together with the property $\hat{P}_S \ket{\psi(t,x)}_{S} = i \frac{\partial}{\partial x} \ket{\psi(t,x)}_{S}$ lead to: 
\begin{equation}\label{evfinale31+1}
i \frac{\partial}{\partial t} \ket{\psi(t, x)}_S \simeq - \frac{1}{2m}\frac{\partial^{2}}{\partial x^{2}}  \ket{\psi(t,x)}_{S}
\end{equation}
which clearly describes the dynamics of the particle in $S$ with respect to the coordinates of the $1+1$ dimensional quantum reference frame. 
We emphasize here that, in this case, we can write the equation (\ref{evfinale31+1}) for the state $\ket{\psi(t,x)}_{S}$ because the values of time and space of the subspaces $C$ and $R$ enter as parameters in $S$ thanks to the entanglement present in the global state $\ket{\Psi}$. We will return to this point later, in Section V, when we also discuss the example of relativistic particles. 

\begin{widetext}

\subsection{A simple Example}
We consider here a simple example assuming $R$ and $S$ as free particles with mass $M$ and $m$ respectively and $d_R=d_S =3$. We start assuming $D_R =D_S = d_R =d_S$, $L_R=L_S=L$ and discrete values of space and time. We have therefore: $p^{(R)}_k = p^{(S)}_k = p_0 + \frac{2\pi}{L} k$,	
with $p_0=-\frac{2\pi}{L}$ (that implies $p_1=0$, $p_2=\frac{2\pi}{L}$), $x^{(R)}_j = x_0+j\frac{L}{3} = 0,\frac{L}{3},\frac{2L}{3}$ and $y^{(S)}_l = y_0+l\frac{L}{3} = 0,\frac{L}{3},\frac{2L}{3}$. The global state satisfying the constraints on total energy and total momentum can be written as:
\begin{equation}
\ket{\Psi} = c_0 \ket{E_{2,0}}_C\ket{p_2}_R\ket{p_0}_S  + c_1 \ket{E_{1,1}}_C\ket{p_1}_R\ket{p_1}_S + c_2 \ket{E_{0,2}}_C\ket{p_0}_R\ket{p_2}_S
\end{equation}
where $E_{k,n} = -( \frac{p^{2}_k}{2M} + \frac{p^{2}_n}{2m} )$ and where we assume, for simplicity, the coefficients $c_i$ to be real. Furthermore we have $E_{2,0}= E_{0,2}=\epsilon= \left(\frac{2\pi}{L}\right)^2( \frac{1}{2M} + \frac{1}{2m})$ and $E_{1,1}=0$. We can now expand the global state $\ket{\Psi}$ simultaneously on the coordinates $\left\{\ket{x_j}_R\right\}$ in $R$ and on the time basis $\left\{\ket{t_m}_C\right\}$ in $C$, and then we search the state $\ket{\psi(t_m,x_j)}_S = \sqrt{d_C}\sqrt{3}(\bra{t_m}\otimes\bra{x_j})\ket{\Psi}$, thus obtaining:
\begin{equation}
\begin{split}
\ket{\psi(t_m,x_j)}_S &= \sqrt{3} \left[c_0 e^{-i\epsilon t_m}\braket{x_j|p_2}\ket{p_0}_S  + c_1 \braket{x_j|p_1}\ket{p_1}_S + c_2  e^{-i\epsilon t_m}\braket{x_j|p_0}\ket{p_2}_S \right] =  \\&
=c_0  e^{-i\epsilon t_m}e^{i\frac{2\pi}{L}x_j}\ket{-\frac{2\pi}{L}}_S  +c_1 \ket{0}_S +c_2 e^{-i\epsilon t_m}e^{-i\frac{2\pi}{L}x_j}\ket{\frac{2\pi}{L}}_S.
\end{split}
\end{equation}
We can now calculate the conditional probability of obtaining $y_l$ on $S$ conditioned on having $x_j$ on $R$ and $t_m$ on $C$. Considering $d_S = D_S =3$, we have:
\begin{equation}\label{esempio1}
P(y_l \: on\: S\:|\:x_j\:on\:R,\: t_m \: on \:C) = \left| \braket{y_l| \psi(t_m,x_j)}\right|^2 =  \frac{1}{3}\left| c_0  e^{-i\epsilon t_m}e^{-i\frac{2\pi}{L}(y_l - x_j)}  + c_1 +  c_2 e^{-i\epsilon t_m}e^{i\frac{2\pi}{L}(y_l-x_j)} \right|^2 .
\end{equation}
Proceeding with the calculations from equation (\ref{esempio1}) and remembering we have real coefficients, we obtain
\begin{multline}
P(y_l \: on\: S\:|\:x_j\:on\:R,\: t_m \: on \:C) = \frac{1}{3} + \frac{2}{3}c_0 c_1 \cos(\epsilon t_m + \frac{2\pi}{L}( y_l - x_j))+ \\ + \frac{2}{3}c_1 c_2 \cos(\epsilon t_m - \frac{2\pi}{L}(y_l -x_j))  + \frac{2}{3}c_0 c_2\left( 1- 2\sin^{2}( \frac{2\pi}{L}(y_l-x_j))  \right) 
\end{multline}
that is the expression of how the probability of having a certain relative distance between the particles $S$ and $R$ given $x_j$ for the reference frame $R$ at time $t_m$. 

If we consider the limiting cases $s,z\longrightarrow\infty$ (mantaining the assumption $L_R=L_S=L$), 
we obtain for the probability density:
\begin{multline}
P(y \: on\: S\:|\:x\:on\:R,\: t \: on \:C) = \frac{1}{L} + \frac{2}{L}c_0 c_1 \cos(\epsilon t + \frac{2\pi}{L}( y - x))+ \\ + \frac{2}{L}c_1 c_2 \cos(\epsilon t - \frac{2\pi}{L}(y -x))  + \frac{2}{L}c_0 c_2\left( 1- 2\sin^{2}( \frac{2\pi}{L}(y-x))  \right) .
\end{multline}

\end{widetext}

\subsection{On the Quantum Speed Limit Time}

An important question is the time scale over which the conditioned state $\ket{\phi (t)}_{R,S}$ evolve into an othogonal configuration, thus becoming fully distinguishable. In dealing with this topic it is useful what was already said in Section II.C. For the sake of simplicity we consider again the case of $d_R=D_R$, $d_S=D_S$ that leads to discrete values of space and orthogonal states. For the temporal degree of freedom we consider the limiting case $s \longrightarrow \infty$. The conditioned state of $R+S$ can be obtained from (\ref{statoglobalespaziotempo}), calculating $\ket{\phi(t)}_{R,S} = \braket{t|\Psi}$. We have ($\hslash \ne 1$):
\begin{equation}\label{equazione}
\ket{\phi(t)}_{R,S} = \sum_{k=0}^{d_S - 1} c_k e^{-i \hslash^{-1} \epsilon_k t}\ket{p=-p_k}_R\otimes\ket{p_k}_S
\end{equation} 
where $\epsilon_k= E^{(R)}(-p_k) + E^{(S)}(p_k)$ is the energy function related to the momenta $p_k$ of $R$ and $S$. Starting from equation (\ref{equazione}) we can calculate in the space $R+S$:

\begin{equation}\label{ultima1}
\begin{split}
\braket{\phi (t_0) | \phi (t)}  = \sum_{k=0}^{d_S - 1} |c_k|^2 e^{-i \hslash^{-1} \epsilon_k (t - t_0)} .
\end{split}
\end{equation} 
Looking at equation (\ref{ultima1}), we can now consider the quantum speed limit time $\delta t$ which gives us the minimum time needed for $R+S$ to evolve to an orthogonal configuration. We have \cite{speedlimit,margolus}:
\begin{equation}\label{dalpha}
\delta t  \geq  max \left( \frac{\pi\hslash}{2 E_{R,S}}  , \frac{\pi\hslash}{2 \Delta E} \right)
\end{equation}
where $E_{R,S} = \bra{\phi(t_0)}\hat{H}_R + \hat{H}_S\ket{\phi(t_0)} $ and $\Delta E$ is the spread in energy related to the coefficients $c_k$ through 
\begin{equation}
\Delta E = \sqrt{\bra{\phi(t_0)} (\hat{H}_R + \hat{H}_S - E_{R,S} )^2\ket{\phi(t_0)}}. 
\end{equation}

The aspect we emphasize is that, as in the case of space, the function $f(t - t_0) = \braket{\phi (t_0) | \phi (t)}$, and consequently the time scale on which $R+S$ varies significantly, is not related to the overlap of the states of the clock. This can be seen considering that in $f(t - t_0)$ do not enter time values different from $t, t_0$ and $f(t - t_0)$ takes the form expressed in Ref. \onlinecite{speedlimit}. So the fact that our time states are not orthogonal does not have a consequence on the speed at which the state $\ket{\phi(t)}_{R,S}$ evolves with respect to $t$. Rather $f(t - t_0)$ is related to the spread of the coefficients $c_k$ appearing in the state (\ref{equazione}). These considerations, together with equation (\ref{dalpha}), indicate the key point in this framework: a large spread in the coefficients $c_k$ within state $\ket{\phi(t)}_{R,S}$, and so in the global state 
$\ket{\Psi} = \sum_{k=0}^{d_S -1} c_k \ket{E=-\epsilon_k}_C\otimes\ket{p=-p_k}_R\otimes\ket{p_k}_S$, is needed to make the time evolution of the subsystem $R+S$ faster \cite{asimmetry}.

\subsection{Introducing Potentials in $R$ and $S$}
We illustrate here the case in which potentials are present within $R$ and $S$ by considering these two subspaces consisting of two harmonic oscillators. We use this example also to extend our framework assuming momentum operators with continuous spectra. So far we have indeed always used momentum operators in $R$ and $S$ with discrete spectra, but the entire discussion can be easily generalized to the case of continuous values for momenta and coordinates, with orthogonal position states. Such a framework allows us to more easily address the problem of considering the presence of potentials in $R$ and $S$. What we need is indeed to have Hermitian operators $\hat{X}$ (within $R$) and $\hat{Y}$ (within $S$) in order to describe the potentials of the harmonic oscillators. In the subspace $C$ we assume the framework described so far working in the limiting case $s \longrightarrow \infty$.

We therefore start considering the Hamiltonians of $R$ and $S$ written as:
\begin{equation}
\begin{split}
& \hat{H}_R= \frac{\hat{P}^{2}_R}{2M} + V_R(\hat{X}) =  \frac{\hat{P}^{(2)}_R}{2M} + \frac{1}{2}M\omega_R\hat{X}^{2}
\\&  \hat{H}_S= \frac{\hat{P}^{2}_S}{2m} + V_S(\hat{Y}) =  \frac{\hat{P}^{(2)}_S}{2m} + \frac{1}{2}m \omega_S \hat{Y}^{2} .
\end{split}
\end{equation} 
In this framework, the global state $\ket{\Psi}$ satisfying the constraint on total momentum $\left( \hat{P}_R + \hat{P}_S\right)\ket{\Psi}=0$ is:
\begin{equation}\label{Gstato1}
\ket{\Psi} = \sum_{n=0}^{d_C - 1} \int dp ~ c_n \psi(p) \ket{E_n}_C \otimes \ket{-p}_R \otimes \ket{p}_S
\end{equation}
where $\sum_{n=0}^{d_C - 1} |c_n|^2 = \int dp |\psi(p)|^2=1$. The state (\ref{Gstato1}) can be rewritten in the energy eigenbasis for $R$ and $S$ as
\begin{multline}\label{Gstato2}
\ket{\Psi} = \sum_{n=0}^{d_C - 1} \sum_{k} \sum_{l} \int dp ~ c_n \psi(p) \\
\times \beta(-p,E_k) \beta(p,E_l) \ket{E_n}_C \otimes \ket{E_k}_R \otimes \ket{E_l}_S
\end{multline}
where $\beta(a,b) = \braket{b|a}$. Through the state (\ref{Gstato2}) we can now impose the constraint on total energy $\hat{H}\ket{\Psi} = \left( \hat{H}_C + \hat{H}_R + \hat{H}_S\right)\ket{\Psi}=0$ where the Hamiltonians of $R$ and $S$ can be rewritten ($\hslash=1$): $\hat{H}_R = \omega_R\left( \hat{a}^{\dagger}_R\hat{a}_R + \frac{1}{2} \right)$ and $\hat{H}_S = \omega_S\left( \hat{a}^{\dagger}_S \hat{a}_S + \frac{1}{2} \right)$, being $\hat{a}^{\dagger}_R$, $\hat{a}^{\dagger}_S$, $\hat{a}_R$ and $\hat{a}_S$ the usual rising and lowering operators for the subsystems $R$ and $S$. For the global state $\ket{\Psi}$ we thus find:
\begin{multline}\label{Gstatoe}
\ket{\Psi} = \sum_{k} \sum_{l} \int dp ~ \tilde{c}_{kl} \psi(p) \\
\times \beta(-p,E_k) \beta(p,E_l) \ket{E=-\epsilon_{kl}}_C \otimes \ket{E_k}_R \otimes \ket{E_l}_S
\end{multline}
with
\begin{equation}
\epsilon_{kl} = \omega_R\left( k + \frac{1}{2} \right) + \omega_S \left( l + \frac{1}{2} \right) .
\end{equation}
The state (\ref{Gstatoe}) provides the explicit form of the global state of our quantum Universe simultaneously satisfying both the constraints: on total energy and total momentum. We can then expand again the subspaces $R$ and $S$ back on the momentum eigenbasis and rewriting (\ref{Gstatoe}) as
\begin{multline}\label{Gstatodef}
\ket{\Psi} = \int dp' \sum_{k} \sum_{l} \int dp ~ \tilde{c}_{kl} \psi(p) \beta(-p,E_k) \beta(p,E_l) \\
\times \beta^{*}(-p',E_k) \beta^{*}(p',E_l) \ket{E=-\epsilon_{kl}}_C \otimes \ket{-p'}_R \otimes \ket{p'}_S .
\end{multline} 

Through the states (\ref{Gstatoe}) and (\ref{Gstatodef}) we can find again all the results shown in Section III.B. For the relative state $\ket{\phi(t)}_{R,S} = \braket{t|\Psi}$ we have: 
\begin{multline}
i \frac{\partial}{\partial t}\ket{\phi(t)}_{R,S}  = \left(\hat{H}_R + \hat{H}_S\right)\ket{\phi(t)}_{R,S} = \\
= \left(\frac{\hat{P}^{(2)}_R}{2M} + \frac{1}{2}M\omega_R\hat{X}^{2} + \frac{\hat{P}^{(2)}_S}{2m} + \frac{1}{2}m \omega_S \hat{Y}^{2}\right)\ket{\phi(t)}_{R,S} .
\end{multline}
which provides the Schrödinger evolution for the subsystem $R+S$. In the same way, for the relative state $\ket{\varphi(x)}_{C,S} = \sqrt{2\pi}\braket{x|\Psi}$ we have: $ \ket{\varphi(x+a)}_{C,S}= e^{-i a \hat{P}_S}\ket{\varphi(x)}_{C,S}$
which shows how the operator $\hat{P}_S$ is the generator of translations in the values of the reference frame $R$ for the relative state of the subsystem $C+S$. Finally we expand the state $\ket{\Psi}$ simultaneously on the coordinates in $R$ and on the time basis in $C$ thus finding:
\begin{equation}\label{Gespansione}
\begin{split}
\ket{\Psi} 
= \frac{1}{T \sqrt{2\pi} } \int dt \int dx \ket{t}_C \otimes \ket{x}_R \otimes \ket{\psi(t,x)}_S
\end{split}
\end{equation}
where the integral on $dt$ is evaluated from $t_0$ and $t_0 +T$ and where now the relative state $\ket{\psi(t,x)}_S =  \sqrt{2\pi} \left( \bra{t}\otimes\bra{x}\right)\ket{\Psi}$ reads
\begin{multline}\label{Gfin}
\ket{\psi(t,x)}_S = \int dp'	\sum_{k} \sum_{l}  \int dp ~ \tilde{c}_{kl} \psi(p) \beta(-p,E_k) \beta(p,E_l) \\ \times \beta^{*}(-p',E_k) \beta^{*}(p',E_l) e^{-i\epsilon_{kl}t} e^{-ixp'} \ket{p'}_S .     
\end{multline} 
Through the state (\ref{Gfin}) 
we can search also in this case the probability density $P(y \: on\: S\:|\:x\:on\:R,\: t \: on \:C) = \left| \braket{y|\psi(t,x)} \right|^2$ obtaining:
\begin{multline}\label{Gprobfinale}
P(y \: on\: S\:|\:x\:on\:R,\: t \: on \:C) =\\= \frac{1}{2\pi} |\int dp' \sum_{k} \sum_{l}  \int dp ~ \tilde{c}_{kl} \psi(p) \beta(-p,E_k) \beta(p,E_l) \\ \times \beta^{*}(-p',E_k) \beta^{*}(p',E_l) e^{-i\epsilon_{kl}t} e^{ip'(y-x)}|^2
\end{multline}
which still depends on time $t$ and on the relative distance $y-x$ between $S$ and $R$.


	\section{Multiple Time Measurements}
Throughout this work, we have seen how conditional probabilities lie at the heart of the PaW mechanism. This aspect of the theory has been criticised by K. V. Kuchar \cite{kuchar} who emphasized that the PaW mechanism is not able to provide the correct propagators when considering multiple measurements. Indeed measurements performed at two times return wrong statistics because the first measurement \lq\lq collapses\rq\rq the time state and freezes the rest of the Universe (namely $R$ and $S$ in our framework). Two possible ways out of this problem have been proposed: the first by R. Gambini, R. A. Porto, J. Pullin, and S. Torterolo (GPPT) in Ref. \onlinecite{gppt} (see also Ref. \onlinecite{esp1}) and the second by V. Giovannetti, S. Lloyd and L. Maccone (GLM) in Ref. \onlinecite{lloydmaccone}. 

We will now explore how these proposals can be applied to our framework of emerging spacetime. For the sake of simplicity we consider again a simple case: we start from the framework described in Sections II.A and III.A, we work with discrete values of space and time and we choose $d_C=D_C$, $d_R=D_R$, $d_S=D_S$. This latter assumption implies that we are considering an equally-spaced spectrum for $\hat{H}_C$ and we emphasize that this choice leads to orthogonal states of time and position, namely  $\braket{t_m|t_{m'}}=\delta_{m,m'}$ on $C$, $\braket{x_i|x_{j}}=\delta_{i,j}$ on $R$ and $\braket{y_l|y_{k}}=\delta_{l,k}$ on $S$. 
In the case of GPPT theory, this simplified framework can then be readily generalized to the case of unequally-spaced levels for $\hat{H}_C$ and to the limiting cases $z,s \longrightarrow \infty$, where time and position are represented by POVMs. Conversely, for the GLM proposal the assumption of orthogonal time states in $C$ and orthogonal position states in $R$ and $S$ will be necessary.

\subsection{The GPPT proposal}
As pointed out in Ref. \onlinecite{esp1} one of the main ingredients of the GPPT theroy is the averaging over the abstract coordinate time (the \lq\lq external time\rq\rq) in order to eliminate any external time dependence in the observables. With the perspective of calculating the probabilities for multiple time measurements, we look in this section at the probability $P(y_l \: on \: S, \: x_j \: on \: R \:  | \: t_m \: on \: C)$ of having $y_l$ on $S$ and $x_j$ on $R$ conditioned to having $t_m$ on $C$. This probability essentially is not different from the probabilities we have calculated so far (apart from a numerical factor). We will see indeed that it will depend on the relative distance between $R$ and $S$ in the same way as we found in the previous Section. What clearly is different is the interpretation of this probability, where the value of the reference frame $R$ is not given and can vary. Following GPPT this probability is given by \cite{gppt}:
\begin{multline}\label{cpGPPT}
P(y_l \: on \: S, \: x_j \: on \: R \:  | \: t_m \: on \: C) = \frac{\int d\theta \: Tr \left[ \hat{\Pi}_{t_m,x_j,y_l}(\theta) \hat{\rho} \right]}{\int d\theta \: Tr \left[ \hat{\Pi}_{t_m}(\theta) \hat{\rho} \right]}=\\
= \frac{1}{d_R} \frac{1}{d_S} \left| \sum_{k=0}^{d_S - 1} c_k e^{-i\epsilon_k t_m } e^{i p_k (y_l - x_j)}\right|^2
\end{multline}
where $\hat{\rho}=\ket{\Psi}\bra{\Psi}= \sum_{k}\sum_{k'}c_k c^{*}_{k'}\ket{E=-\epsilon_k}\bra{E=-\epsilon_{k'}}\otimes\ket{p=-p_k}\bra{p=-p_{k'}}\otimes\ket{p_k}\bra{p_{k'}}$ is the global state of the Universe, $\theta$ is the external time, $\hat{\Pi}_{t_m}(\theta)=\hat{U}^{\dagger} (\theta) \hat{\Pi}_{t_m} \hat{U}(\theta)$ (with $\hat{U}(\theta) = e^{-i\hat{H}\theta}$) is the projector relative to the result $t_m$ for a clock measurement at external time $\theta$ and $\hat{\Pi}_{t_m,x_j,y_l}(\theta)=\hat{U}^{\dagger} (\theta) \hat{\Pi}_{t_m,x_j,y_l} \hat{U}(\theta)$ is the projector relative to the result $y_l$ for a measurement on $S$, $x_j$ for a measurement on $R$ and $t_m$ for a measurement on $C$ at external time $\theta$ (we are working here in the Heisenberg picture with respect to the external time $\theta$). Equation (\ref{cpGPPT}) now takes the place of equations (\ref{probfinalediscreta}) and (\ref{probfinale}) and, as expected, depends on time $t_m$ and on the distance $y_l - x_j$ between the two subsystems $S$ and $R$. 

Equation (\ref{cpGPPT}) can be generalized to the case of multiple time measurements. For two measurements at times $t_m$ and $t_{m'}$ (with $t_{m'} > t_m$) we have: 
\begin{multline}\label{cpGPPT2}
P(y_{l'} \: on \: S, x_{j'} \: on \: R \: | \:t_{m'} \: on \: C,y_l, x_j, t_m) =\\
= \frac{\int d\theta \int d\theta' \: Tr \left[\hat{\Pi}_{t_{m'},x_{j'},y_{l'}}(\theta)  \hat{\Pi}_{t_m,x_j,y_l}(\theta') \hat{\rho} \hat{\Pi}_{t_m,x_j,y_l}(\theta') \right]}{\int d\theta \int d\theta' \: Tr \left[\hat{\Pi}_{t_{m'}}(\theta) \hat{\Pi}_{t_m,x_j,y_l}(\theta') \hat{\rho} \hat{\Pi}_{t_m,x_j,y_l}(\theta') \right]}
\end{multline}
which provides the conditional probability of obtaining $y_{l'}$ and $x_{j'}$ on $S$ and $R$ at clock time $t_{m'}$, given that a \lq\lq previous\rq\rq joint measurement of $S$, $R$ and $C$ returns $y_l$, $x_j$ and $t_m$. Proceeding with the calculations from the equation (\ref{cpGPPT2}) we obtain:
\begin{multline}\label{cpGPPT3}
P(y_{l'} \: on \: S, x_{j'} \: on \: R \: | \:t_{m'} \: on \: C,y_l, x_j, t_m) =\\
= \frac{1}{d^{2}_R} \frac{1}{d^{2}_S} \left| \sum_{k=0}^{d_S-1} e^{-i\epsilon_k(t_{m'} - t_m)} e^{ip_k(\Delta_f - \Delta_i)}\right|^2
\end{multline}
where we have written $\Delta_i=y_l - x_j$ and $\Delta_f=y_{l'} - x_{j'}$ respectively the initial distance and the final distance (namely the distance at the first measurement at time $t_m$ and the distance at the second measurement at time $t_{m'}$) between the particle $S$ and $R$. This equation provides the the correct propagator (the same result can be indeed obtained calculating $\left|\bra{x_{j'}} \bra{y_{l'}} e^{-i(\hat{H}_R + \hat{H}_S)(t_{m'} - t_m)}\ket{x_j}\ket{y_l}\right|^2$) and again depends on the initial and the final relative distances between $S$ and $R$. 

As previously mentioned, these results can be readily generalized to the case of unequally-spaced levels for $\hat{H}_C$ and to the limiting cases $z,s \longrightarrow \infty$. Indeed, in our framework, the fact of using POVMs in describing time and space does not constitute a problem in the application of the GPPT theory because when we calculate the probability $P(y_{l'} \: on \: S, x_{j'} \: on \: R \: | \:t_{m'} \: on \: C,y_l, x_j, t_m)$ through the (\ref{cpGPPT2}) we do not find terms related to the overlap between the states and therefore interference phenomena are not present even if the states are not orthogonal \cite{nostro2}.

\subsection{GLM's Multiple Measurements}
We focus now on the GLM proposal applying it directly to our case of emergent spacetime. In this Section we consider the global state of the Universe written as in (\ref{serveperGLM}), that in our particular case of orthogonal states of time and space ($d_C=D_C$, $d_R=D_R$, $d_S=D_S$) becomes

\begin{equation}\label{statoGLM}
\begin{split}
\ket{\Psi} 
= \frac{1}{\sqrt{d_C}}\sum_{m=0}^{d_C -1} \ket{t_m}_C \otimes \hat{U}_{R,S}(t_m - t_0)\ket{\phi(t_0)}_{R,S}
\end{split}
\end{equation}
where $\ket{\phi(t_0)}_{R,S}$ is the state of $R+S$ conditioned on $t_0$ that is the value of the clock taken as initial time and where thanks to the (\ref{evoluzioneRS}) we have defined $\hat{U}_{R,S}(t_m - t_0)=e^{-i(\hat{H}_R + \hat{H}_S)(t_{m} - t_0)}$.

Again we start considering to perform a single measurement within $R+S$ at time $t_{m'}$. We divide the spaces $R$ and $S$ into two subsystems respectively: $\mathcal{H}_R = \mathcal{H}_{Q_R}\otimes\mathcal{H}_{M_R}$ and $\mathcal{H}_S = \mathcal{H}_{Q_S}\otimes\mathcal{H}_{M_S}$ where $Q_R$, $Q_S$ are the systems to be measured (the \textit{observed}) and $M_R$, $M_S$ are the ancillary memory systems (the \textit{observers}). In this framework GLM use von Neumann's prescription for measurements \cite{vonneumann}, where a measurement apparatus essentially consists in an (ideally instantaneous) interaction between the observed and the observers. The interaction correlates $Q_R$ with $M_R$ and $Q_S$ with $M_S$ along the eigenbasis $\{ \ket{x_j , y_l}\}$ of the observables $\hat{X}=\sum_{j}x_j\ket{x_j}\bra{x_j}$ and $\hat{Y}= \sum_{l}y_l\ket{y_l}\bra{y_l}$ to be measured, that is  
\begin{widetext}
\begin{equation}\label{mapping}
\ket{\phi(t_m)}_{Q_R,Q_S}\otimes\ket{r,r}_{M_R,M_S} \rightarrow \sum_{j=0}^{d_R -1}\sum_{l=0}^{d_S -1}\left( \braket{x_j,y_l|\phi(t_m)} \right)\ket{x_j,y_l}_{Q_R,Q_S}\otimes\ket{x_j,y_l}_{M_R,M_S}
\end{equation}
where $\ket{r,r}_{M_R,M_S}$ is the stete of the memories before the interaction and $\braket{x_j,y_l|\phi(t_m)}$ is the probabiliy amplitude of obtaining $x_j$ and $y_l$ when measuring the observables $\hat{X}$ and $\hat{Y}$. In this framework the Hamiltonian of $R+S$ can be written as
\begin{equation}\label{hamiltonianaGLM}
\hat{H}_R(t_m) + \hat{H}_S(t_m)= \hat{H}_{Q_R} + \hat{H}_{Q_S} + \delta_{m,m'}\left( \hat{h}_{Q_R,M_R} + \hat{h}_{Q_S,M_S} \right)
\end{equation} 
where $\hat{h}_{Q_i,M_i}$ are responsible for the mapping equation (\ref{mapping}). So we can write the global state (\ref{statoGLM}) including the measurement thus obtaining \cite{lloydmaccone,esp2}
\begin{multline}\label{singolamisuraGLM}
\ket{\Psi} = \frac{1}{\sqrt{d_C}}\sum_{m < m'} \ket{t_m}_C \otimes \hat{U}_{R,S}(t_m - t_0)\ket{\phi(t_0)}_{Q_R,Q_S}\otimes\ket{r,r}_{M_R,M_S} + \\
+ \frac{1}{\sqrt{d_C}}\sum_{m \ge m'} \ket{t_m}_C \otimes \sum_{j=0}^{d_R -1}\sum_{l=0}^{d_S -1}\left( \braket{x_j,y_l|\phi(t_m)} \right) \hat{U}_{Q_R,Q_S}(t_m - t_{m'})\ket{x_j,y_l}_{Q_R,Q_S}\otimes\ket{x_j,y_l}_{M_R,M_S} 
\end{multline}
where the first summation describes the evolution of $R+S$ prior to the measurement, when the memories are in the state $\ket{r,r}_{M_R,M_S}$, whereas the second summation describes the evolution after the measurement, when the memories are correlated with the subsystems $Q_R$ and $Q_S$. Now the probability that, at a given time $t_{m'}$, the values $x_{j'}$ and $y_{l'}$ will be registered by the memories $M_R$ and $M_S$ respectively can be expressed as \cite{lloydmaccone}:
\begin{equation}\label{probGLM}
P(x_{j'},y_{l'} \:  | \: t_{m'} ) = \frac{|| \left( _{C}\bra{t_{m'}}\otimes _{M_R}\bra{x_{j'}}\otimes _{M_S}\bra{y_{l'}}\right)\ket{\Psi}||^2}{1/d_C}
\end{equation}
where we use the norm of a vector as $||\ket{v}||^2 = \braket{v|v}$. Equation (\ref{probGLM}) returns the correct result 
\begin{equation}
P(x_{j'},y_{l'} \:  | \: t_{m'} ) =  \left| \braket{x_{j'},y_{l'}|\phi(t_{m'})} \right|^2 = \frac{1}{d_R} \frac{1}{d_S} \left| \sum_{k=0}^{d_S - 1} c_k e^{-i\epsilon_k t_{m'} } e^{i p_k (y_{l'} - x_{j'})}\right|^2  . 
\end{equation}
The GLM formalism allows us easily to calculate also the probability $P(y_{l'} \:  | \:x_{j'}, t_{m'} )$ simply by dividing equation (\ref{probGLM}) by $1/d_R$. In this case we obtain: $P(y_{l'} \:  | \:x_{j'}, t_{m'} ) = \frac{1}{d_S} \left| \sum_{k=0}^{d_S - 1} c_k e^{-i\epsilon_k t_{m'} } e^{i p_k (y_{l'} - x_{j'})}\right|^2$ in accordance with our previous results.

It is now possible to extend equations (\ref{hamiltonianaGLM}), (\ref{singolamisuraGLM}) and (\ref{probGLM}) in order to perform multiple measurements. The framework allows an arbitrary number of measurements but we consider here only the simple case of a double measurement at times $t_{m'}$ and $t_{m''}$ (with $t_{m''} > t_{m'}$). What we have to do is simply consider a larger set of memories $M^{(1)}_R$, $M^{(1)}_S$, $M^{(2)}_R$ and $M^{(2)}_S$ (where $M^{(1)}_R$, $M^{(1)}_S$ refer to the first measurement and $M^{(2)}_R$, $M^{(2)}_S$ refer to the second measurement) which couple with $Q_R$ and $Q_S$ through the time-dependent Hamiltonian
\begin{equation}
\hat{H}_R(t_m) + \hat{H}_S(t_m)= \hat{H}_{Q_R} + \hat{H}_{Q_S} + \delta_{m,m'}\left( \hat{h}_{Q_R,M^{(1)}_R} + \hat{h}_{Q_S,M^{(1)}_S} \right) 
+ \delta_{m,m''}\left( \hat{h}_{Q_R,M^{(2)}_R} + \hat{h}_{Q_S,M^{(2)}_S} \right) .
\end{equation}
The global state (\ref{statoGLM}) including the double measurement can then be written \cite{lloydmaccone,esp2}
\begin{multline}\label{doppiamisuraGLM}
\ket{\Psi} = \frac{1}{\sqrt{d_C}}\sum_{m < m'} \ket{t_m}_C \otimes \hat{U}_{R,S}(t_m - t_0)\ket{\phi(t_0)}_{Q_R,Q_S}\otimes\ket{r,r}_{M^{(1)}_R,M^{(1)}_S}\otimes \ket{r,r}_{M^{(2)}_R,M^{(2)}_S} + \\
+ \frac{1}{\sqrt{d_C}} \sum_{m' \le m < m''} \ket{t_m}_C \otimes \sum_{j=0}^{d_R -1}\sum_{l=0}^{d_S -1}\left( \braket{x_j,y_l|\phi(t_m)} \right)\\ 
\times \hat{U}_{Q_R,Q_S}(t_m - t_{m'})\ket{x_j,y_l}_{Q_R,Q_S}  \otimes\ket{x_j,y_l}_{M^{(1)}_R,M^{(1)}_S}\otimes\ket{r,r}_{M^{(2)}_R,M^{(2)}_S} + \\
+ \frac{1}{\sqrt{d_C}} \sum_{m \ge m''} \ket{t_m}_C \otimes  \sum_{i=0}^{d_R -1}\sum_{k=0}^{d_S -1}  \sum_{j=0}^{d_R -1}\sum_{l=0}^{d_S -1} \left( \bra{x_i,y_k}\hat{U}_{Q_R,Q_S}(t_{m''} -t_{m'})\ket{x_j,y_l}\right) \left( \braket{x_j,y_l|\phi(t_m)}\right) \\
\times \hat{U}_{Q_R,Q_S}(t_m - t_{m''})\ket{x_i,y_k}_{Q_R,Q_S}\otimes\ket{x_j,y_l}_{M^{(1)}_R,M^{(1)}_S}\otimes\ket{x_i,y_k}_{M^{(2)}_R,M^{(2)}_S} .
\end{multline}
Through the state (\ref{doppiamisuraGLM}) we search now the probability of obtaining $x_{j''}$ and $y_{l''}$ on $R$ and $S$ at time $t_{m''}$, given that a \lq\lq previous\rq\rq measurement at time $t_{m'}$ returns $x_{j'}$ and $y_{l'}$. This can be formally expressed as follows \cite{lloydmaccone}
\begin{multline}\label{probdoppioGLM}
P\left(    \left( x_{j''},y_{l''} \:  | \: t_{m''} \right)\:|\:\left( x_{j'},y_{l'} \:  | \: t_{m'} \right)  \right) = \frac{P(x_{j''},y_{l''},x_{j'},y_{l'}\:|\:t_{m''})}{P(x_{j'},y_{l'} \:  | \: t_{m'})}=\\
= \frac{||(_{C}\bra{t_{m''}}\otimes _{M^{(1)}_R}\bra{x_{j'}}\otimes _{M^{(1)}_S}\bra{y_{l'}} \otimes _{M^{(2)}_R}\bra{x_{j''}}\otimes _{M^{(2)}_S}\bra{y_{l''}})\ket{\Psi} ||^2}{||  (_{C}\bra{t_{m'}}\otimes _{M^{(1)}_R}\bra{x_{j'}}\otimes _{M^{(1)}_S}\bra{y_{l'}})\ket{\Psi} ||^2} 
\end{multline}
which returns the correct result for a two-times measurement:
\begin{multline}\label{pf} 
P\left(    \left( x_{j''},y_{l''} \:  | \: t_{m''} \right)\:|\:\left( x_{j'},y_{l'} \:  | \: t_{m'} \right)  \right) =\\= \left| \bra{x_{j''},y_{l''}}\hat{U}_{Q_R,Q_S}(t_{m''} -t_{m'})\ket{x_{j'},y_{l'}}\right|^2 
= \frac{1}{d^{2}_R} \frac{1}{d^{2}_S} \left| \sum_{k=0}^{d_S-1} e^{-i\epsilon_k(t_{m''} - t_{m'})} e^{ip_k(\Delta_f - \Delta_i)}\right|^2
\end{multline}
where $\Delta_i=y_{l'} - x_{j'}$ and $\Delta_f=y_{l''} - x_{j''}$ are again the distances between $R$ and $S$ at times $t_{m'}$ and $t_{m''}$ respectively.
The result (\ref{pf}) follows from the fact that 
\begin{equation}
P(x_{j''},y_{l''},x_{j'},y_{l'}\:|\:t_{m''}) = | \left( \bra{x_{j''},y_{l''}}\hat{U}_{Q_R,Q_S}(t_{m''} -t_{m'})\ket{x_{j'},y_{l'}}\right)  \times \left( \braket{x_{j'},y_{l'}|\phi(t_{m'})}\right) |^2
\end{equation}
obtained from the third summation in (\ref{doppiamisuraGLM}) and $P(x_{j'},y_{l'} \:  | \: t_{m'}) = \left| \braket{x_{j'},y_{l'}|\phi(t_{m'})} \right|^2$ obtained from the second summation in (\ref{doppiamisuraGLM}). It is finally easy to verify that the probability $P\left(    \left( y_{l''} \:  | \: x_{j''}, t_{m''} \right)\:|\:\left( y_{l'} \:  | \:x_{j'}, t_{m'} \right)  \right)$ can be obtained dividing (\ref{pf}) by $1/d^2_{R}$. So we find: 
\begin{equation}
	P\left(    \left( y_{l''} \:  | \: x_{j''}, t_{m''} \right)\:|\:\left( y_{l'} \:  | \:x_{j'}, t_{m'} \right)  \right)=\frac{1}{d^{2}_S} \left| \sum_{k=0}^{d_S-1} e^{-i\epsilon_k(t_{m''} - t_{m'})} e^{ip_k(\Delta_f - \Delta_i)}\right|^2.
\end{equation}

As we mentioned previously the GLM proposal can not be generalized, in our framework, to the case of non-orthogonal states of time and space and consequently in applying the GLM theory we are forced to assume $D_R=d_R$, $D_S=d_S$ and equally-spaced energy levels in $\hat{H}_C$.


	\section{Generalization to $3+1$ Dimensional Spacetime}
\label{Generalization}
We generalize here the results of Section III.B to the case of $3+1$ dimensional spacetime, meaning that we have now three degrees of freedom within the subspaces $R$ and $S$. The constraint on total energy reads again:

\begin{equation}\label{4constraint1}
\hat{H}\ket{\Psi} = ( \hat{H}_C + \hat{H}_R +\hat{H}_S)\ket{\Psi}=0 .
\end{equation}
The Hamiltonians $\hat{H}_R$ and $\hat{H}_S$ in (\ref{4constraint1}) depend respectively on the operators $\hat{P}^{(1)}_R$, $\hat{P}^{(2)}_R$, $\hat{P}^{(3)}_R$ and $\hat{P}^{(1)}_S$, $\hat{P}^{(2)}_S$, $\hat{P}^{(3)}_S$ where $1,2,3$ are the three degrees of freedom identifying three orthogonal directions in space (also in this case, for simplicity, we assume that no potentials are present in $R$ and $S$). The constraint on the total momentum reads now 

\begin{equation}
	\vec{P}\ket{\Psi} = (\vec{P}_R + \vec{P}_S)\ket{\Psi}=0,
\end{equation}
where we consider

\begin{equation}\label{4constraint2}
\begin{split}
  (\hat{P}^{(J)}_R + \hat{P}^{(J)}_S)\ket{\Psi}=0 \:\:\:\:\:\: for \:\: J=1,2,3 .
\end{split}
\end{equation} 
Also in this case, for simplicity we have chosen $\vec{P}_C = 0$ (with $\hat{P}^{(1)}_C = \hat{P}^{(2)}_C=\hat{P}^{(3)}_C=0$). So, assuming $d_C, d^{(1)}_R,d^{(2)}_R,d^{(3)}_R \gg d^{(1)}_S,d^{(2)}_S,d^{(3)}_S$ and simplifying the notation as much as possible, the global state $\ket{\Psi}$ satisfying the constraints (\ref{4constraint1}) and (\ref{4constraint2}) can be written as

\begin{equation}\label{4statoglobale}
\ket{\Psi} = \sum_{i=0}^{d^{(1)}_S - 1} \sum_{j=0}^{d^{(2)}_S - 1} \sum_{k=0}^{d^{(3)}_S - 1} c_{ijk}\ket{-\epsilon_{ijk}}_C \otimes \ket{-p^{(1)}_i , -p^{(2)}_j, -p^{(3)}_k }_R \otimes \ket{p^{(1)}_i , p^{(2)}_j, p^{(3)}_k }_S 
\end{equation}
where $d^{(1)}_R,d^{(2)}_R,d^{(3)}_R$ and $d^{(1)}_S,d^{(2)}_S,d^{(3)}_S$ are the dimensions of the subspaces of $R$ and $S$ (associated with the three spatial directions 1,2,3) and $\epsilon_{ijk}$ is the energy function related with the momenta of $R$ and $S$. In the next two paragraphs we will study separately the case of discrete values for space and time and then the case of continuous values for space and time. 

\subsection{Discrete Values for Space and Time}
Starting from the state $\ket{\Psi}$ satisfying (\ref{4constraint1}) and (\ref{4constraint2}), we can now expand it on the basis $\left\{\ket{t_a}_C\right\}$ in $C$ thanks to (\ref{pomidentity2}), thus obtaining

\begin{equation}
\begin{split}
\ket{\Psi} = \frac{d_C}{D_C} \sum_{a=0}^{D_C -1}\ket{t_a}\braket{t_a|\Psi} = \frac{\sqrt{d_C}}{D_C}\sum_{a=0}^{D_C -1}\ket{t_a}_C \otimes \ket{\phi(t_a)}_{R,S}
\end{split}
\end{equation}
where the relative state $\ket{\phi(t_a)}_{R,S} = \sqrt{d_C}\braket{t_a|\Psi}$ takes now the form

\begin{equation}
\ket{\phi(t_a)}_{R,S} = \sum_{i=0}^{d^{(1)}_S - 1} \sum_{j=0}^{d^{(2)}_S - 1} \sum_{k=0}^{d^{(3)}_S - 1} c_{ijk} e^{-it_a \epsilon_{ijk}} \ket{-p^{(1)}_i , -p^{(2)}_j, -p^{(3)}_k }_R \otimes \ket{p^{(1)}_i , p^{(2)}_j, p^{(3)}_k }_S .
\end{equation}
For this relative state we again easily find the Schrödiger evolution with respect to the clock values, namely

\begin{equation}
\ket{\phi(t_a)}_{R,S} = e^{-i (\hat{H}_R + \hat{H}_S)(t_a - t_0)}\ket{\phi(t_0)}_{R,S}
\end{equation}
where $\ket{\phi(t_0)}_{R,S}= \sqrt{d_C} \braket{t_0|\Psi}$ is the state of $R+S$ conditioned on $t_0$ that is the value of the clock taken as initial time. All the consideration made in Section III.B are clearly still valid in this case and we do not repeat them.

We emphasize here that in each subspace of $R$, related to the three degrees of freedom, we apply the formalism described in Section II. This means considering the states $\ket{x^{(J)}_n} = 1/\sqrt{d^{(J)}_R}\sum_{k=0}^{d^{(J)}_R -1}e^{-i p^{(J)}_k x^{(J)}_n}\ket{p^{(J)}_k}$ and the values $x^{(J)}_n = x^{(J)}_0 + n L^{(J)}_R/D^{(J)}_R$ with $n=0,1,2,...,z=D^{(J)}_R -1$ for $J=1,2,3$. The same clearly holds for the subspaces within the system $S$ that we equip with the states $\ket{y^{(J)}_q} = 1/\sqrt{d^{(J)}_S}\sum_{k=0}^{d^{(J)}_S -1}e^{-i p^{(J)}_k y^{(J)}_q}\ket{p^{(J)}_k}$ and with the values $y^{(J)}_q = y^{(J)}_0 + q L^{(J)}_S/D^{(J)}_S$ with $q=0,1,2,...,z=D^{(J)}_S -1$ (for semplicity we choose $D^{(J)}_R = D^{(J)}_S =z+1 ~ \forall J$). We can now expand the state $\ket{\Psi}$ in the coordinates $\left\{ \ket{x^{(1)}_l , x^{(2)}_m, x^{(3)}_n}_R\right\}$ in $R$:

\begin{equation}
\begin{split}
\ket{\Psi} &=\frac{d^{(1)}_R}{D^{(1)}_R} \frac{d^{(2)}_R}{D^{(2)}_R} \frac{d^{(3)}_R}{D^{(3)}_R} \sum_{l=0}^{D^{(1)}_R - 1} \sum_{m=0}^{D^{(2)}_R - 1} \sum_{n=0}^{D^{(3)}_R - 1}\ket{x^{(1)}_l , x^{(2)}_m, x^{(3)}_n}\braket{x^{(1)}_l , x^{(2)}_m, x^{(3)}_n|\Psi} =\\& 
= \frac{\sqrt{d^{(1)}_R}}{D^{(1)}_R} \frac{\sqrt{d^{(2)}_R}}{D^{(2)}_R} \frac{\sqrt{d^{(3)}_R}}{D^{(3)}_R} \sum_{l=0}^{D^{(1)}_R - 1} \sum_{m=0}^{D^{(2)}_R - 1} \sum_{n=0}^{D^{(3)}_R - 1}\ket{x^{(1)}_l , x^{(2)}_m, x^{(3)}_n}_R \otimes \ket{\varphi(x^{(1)}_l , x^{(2)}_m, x^{(3)}_n)}_{C,S} 
\end{split}
\end{equation}
where $\ket{\varphi(x^{(1)}_l , x^{(2)}_m, x^{(3)}_n)}_{C,S}= \sqrt{d^{(1)}_R d^{(2)}_R d^{(3)}_R} \braket{x^{(1)}_l , x^{(2)}_m, x^{(3)}_n|\Psi}$ is the relative state of $C+S$ conditioned to the value $(x^{(1)}_l , x^{(2)}_m, x^{(3)}_n)$ of the reference frame $R$. For this relative state we find:

\begin{multline}\label{4trasl}
\ket{\varphi(x^{(1)}_l , x^{(2)}_m, x^{(3)}_n)}_{C,S} = \sqrt{d^{(1)}_R d^{(2)}_R d^{(3)}_R} \braket{x^{(1)}_l , x^{(2)}_m, x^{(3)}_n|\Psi}=\\
= \sqrt{d^{(1)}_R d^{(2)}_R d^{(3)}_R} \bra{x^{(1)}_l , x^{(2)}_m, x^{(3)}_n}e^{i\hat{P}^{(1)}_R(x^{(1)}_l - x^{(1)}_0)}e^{i\hat{P}^{(2)}_R(x^{(2)}_m - x^{(2)}_0)}e^{i\hat{P}^{(3)}_R(x^{(3)}_n - x^{(3)}_0)}\ket{\Psi}=\\
=e^{-i\hat{P}^{(1)}_S(x^{(1)}_l - x^{(1)}_0)}e^{-i\hat{P}^{(2)}_S(x^{(2)}_m - x^{(2)}_0)}e^{-i\hat{P}^{(3)}_S(x^{(3)}_n - x^{(3)}_0)} \ket{\varphi(x^{(1)}_0 , x^{(2)}_0, x^{(3)}_0)}_{C,S}
\end{multline}
where we used the relative state definition and the constraint (\ref{4constraint2}). Equation (\ref{4trasl}) can be rewritten in a more compact form as

\begin{equation}
\ket{\varphi(\vec{x_0} + \vec{a})}_{C,S} = e^{-i\vec{a} \cdot \vec{P}_S} \ket{\varphi(\vec{x_0})}_{C,S}
\end{equation}
where $\vec{P}_S = (\hat{P}^{(1)}_S,\hat{P}^{(2)}_S,\hat{P}^{(3)}_S)$, $\vec{x_0} = (x^{(1)}_0 , x^{(2)}_0, x^{(3)}_0)$ is the initial position of the reference frame and the translation vector is $\vec{a} = (x^{(1)}_l - x^{(1)}_0, x^{(2)}_m - x^{(2)}_0, x^{(3)}_n - x^{(3)}_0)$.

Finally we can expand the state $\ket{\Psi}$ simultaneously on time $\left\{\ket{t_a}_C\right\}$ and on the coordinates $\left\{ \ket{x^{(1)}_l , x^{(2)}_m, x^{(3)}_n}_R\right\}$, thus obtaining:

\begin{equation}
\begin{split}
\ket{\Psi} &= \left( \frac{d_C}{D_C} \sum_{a}
\ket{t_a}\bra{t_a} \otimes \frac{d^{(1)}_R}{D^{(1)}_R} \frac{d^{(2)}_R}{D^{(2)}_R} \frac{d^{(3)}_R}{D^{(3)}_R} \sum_{l,m,n} 
\ket{x^{(1)}_l , x^{(2)}_m, x^{(3)}_n}\bra{x^{(1)}_l , x^{(2)}_m, x^{(3)}_n} \right)\ket{\Psi} = \\&
= \frac{\sqrt{d_C}}{D_C}\frac{\sqrt{d^{(1)}_R}}{D^{(1)}_R} \frac{\sqrt{d^{(2)}_R}}{D^{(2)}_R} \frac{\sqrt{d^{(3)}_R}}{D^{(3)}_R} \sum_{a}\sum_{l,m,n} 
\ket{t_a}_C \otimes \ket{x^{(1)}_l , x^{(2)}_m, x^{(3)}_n}_R \otimes \ket{\psi(t_a,x^{(1)}_l , x^{(2)}_m, x^{(3)}_n)}_S 
\end{split}
\end{equation}
where the summation on time runs from $0$ to $D_C -1$, the summations on $l,m,n$ run from $0$ to $D^{(1)}_R -1$, $D^{(2)}_R -1$, $D^{(3)}_R -1$ respectively and where $\ket{\psi(t_a,x^{(1)}_l , x^{(2)}_m, x^{(3)}_n)}_S = \sqrt{d_C}\sqrt{d^{(1)}_R d^{(2)}_R d^{(3)}_R}(\bra{t_a}\otimes\bra{x^{(1)}_l , x^{(2)}_m, x^{(3)}_n})\ket{\Psi}$ is the relative state of the system $S$ at time $t_a$ and conditioned on the value $(x^{(1)}_l , x^{(2)}_m, x^{(3)}_n)$ for the spatial reference frame $R$. Through this state, extending the formalism of Section III.B, we can search the conditional probability:

\begin{multline}\label{4probdiscreta}
P(y^{(1)}_p, y^{(2)}_q,y^{(3)}_r\:|\:x^{(1)}_l , x^{(2)}_m, x^{(3)}_n,t_a) 
=  \frac{d^{(1)}_S}{D^{(1)}_S} \frac{d^{(2)}_S}{D^{(2)}_S} \frac{d^{(3)}_S}{D^{(3)}_S} 
\left|\braket{y^{(1)}_p, y^{(2)}_q,y^{(3)}_r|\psi(t_a,x^{(1)}_l , x^{(2)}_m, x^{(3)}_n)}\right|^2=\\
=\frac{1}{D^{(1)}_S}  \frac{1}{D^{(2)}_S} \frac{1}{D^{(3)}_S} \left| \sum_{i,j,k}
c_{ijk} e^{-it_a\epsilon_{ijk}} e^{ip^{(1)}_i(y^{(1)}_p  -  x^{(1)}_l )} e^{ip^{(2)}_j (y^{(2)}_q  -  x^{(2)}_m )} e^{ip^{(3)}_k(y^{(3)}_r  -  x^{(3)}_n )}\right|^2
\end{multline}  
where the summations on $i,j,k$ run from $0$ to $d^{(1)}_S - 1$, $d^{(2)}_S - 1$ and $d^{(3)}_S - 1$ respectively. Equation (\ref{4probdiscreta}) provides the probability of having a certain position $(y^{(1)}_p, y^{(2)}_q,y^{(3)}_r)$ on $S$ at time $t_a$ and knowing that the spatial reference frame $R$ is in $(x^{(1)}_l , x^{(2)}_m, x^{(3)}_n)$. This conditional probability is well-defined (indeed it is easy to verify that we have $\sum_{p=0}^{D^{(1)}_S - 1} \sum_{q=0}^{D^{(2)}_S - 1} \sum_{r=0}^{D^{(3)}_S - 1} P(y^{(1)}_p, y^{(2)}_q,y^{(3)}_r\:|\:x^{(1)}_l , x^{(2)}_m, x^{(3)}_n,t_a)=1$) and, as expected depends on the relative distance between $R$ and $S$.

\subsection{Continuous Values for Space and Time}
We consider now the case of continuous values for space and time. We therefore assume $s\longrightarrow \infty$ for the space $C$ and $z\longrightarrow \infty$ within the subspaces of $R$ and $S$. This implies that we have the states $\ket{x^{(J)}} = \sum_{k=0}^{d^{(J)}_R -1}e^{-i p^{(J)}_k x^{(J)}}\ket{p^{(J)}_k}$ with $x^{(J)} \in \left[ x^{(J)}_0 , x^{(J)}_0 + L^{(J)}_R \right]$ within $R$ and the states $\ket{y^{(J)}} = \sum_{k=0}^{d^{(J)}_S -1}e^{-i p^{(J)}_k y^{(J)}}\ket{p^{(J)}_k}$ with $y^{(J)} \in \left[ y^{(J)}_0 , y^{(J)}_0 + L^{(J)}_S \right]$ within $S$ for $J=1,2,3$. Starting from the state (\ref{4statoglobale}) we can write:

\begin{equation}
\ket{\Psi} = \frac{1}{T} \int_{t_0}^{t_0 + T} dt \ket{t}\braket{t|\Psi} = \frac{1}{T} \int_{t_0}^{t_0 + T} dt \ket{t}_C\otimes\ket{\phi(t)}_{R,S}
\end{equation}
where the relative state $\ket{\phi(t)}_{R,S} = \braket{t|\Psi}$ takes now the form

\begin{equation}\label{86}
\ket{\phi(t)}_{R,S} = \sum_{i=0}^{d^{(1)}_S - 1} \sum_{j=0}^{d^{(2)}_S - 1} \sum_{k=0}^{d^{(3)}_S - 1} c_{ijk} e^{-it \epsilon_{ijk}} \ket{-p^{(1)}_i , -p^{(2)}_j, -p^{(3)}_k}_R \otimes \ket{p^{(1)}_i , p^{(2)}_j, p^{(3)}_k }_S .
\end{equation}
As in the $1+1$ dimensional case, using (\ref{4constraint1}) and (\ref{86}), for this relative state we find:

\begin{equation}\label{4evcont}
i \frac{\partial}{\partial t}\ket{\phi(t)}_{R,S} = (\hat{H}_R + \hat{H}_S)\ket{\phi(t)}_{R,S}
\end{equation}
which provides the Scrhödinger evolution of $R+S$ with respect to the clock time $t$.   

We can now expand the global state $\ket{\Psi}$ in the coordinate basis $\left\{ \ket{x^{(1)} , x^{(2)}, x^{(3)} }_R\right\}$ in $R$, that is

\begin{equation}\label{87}
\begin{split}
\ket{\Psi} &= \frac{1}{L^{(1)}_R} \frac{1}{L^{(2)}_R} \frac{1}{L^{(3)}_R} \int dx^{(1)} \int dx^{(2)} \int dx^{(3)} \ket{x^{(1)} , x^{(2)}, x^{(3)}} \braket{x^{(1)} , x^{(2)}, x^{(3)}|\Psi}=\\&
=\frac{1}{L^{(1)}_R} \frac{1}{L^{(2)}_R} \frac{1}{L^{(3)}_R} \int dx^{(1)} \int dx^{(2)} \int dx^{(3)} \ket{x^{(1)} , x^{(2)}, x^{(3)}}_R \otimes \ket{\varphi(x^{(1)} , x^{(2)}, x^{(3)})}_{C,S}
\end{split}
\end{equation}
where each integral on $dx^{(J)}$ is evaluated from $x^{(J)}_0$ to $x^{(J)}_0 + L^{(J)}_R$ and where the state $\ket{\varphi(x^{(1)} , x^{(2)}, x^{(3)})}_{C,S}$ takes the form:

\begin{equation}\label{4statoCS}
\ket{\varphi(x^{(1)} , x^{(2)}, x^{(3)})}_{C,S} = \sum_{i,j,k} 
c_{ijk} e^{-ip^{(1)}_i x^{(1)}} e^{-ip^{(2)}_j x^{(2)}} e^{-ip^{(3)}_k x^{(3)}} \ket{-\epsilon_{ijk}}_C \otimes \ket{p^{(1)}_i , p^{(2)}_j, p^{(3)}_k }_S 
\end{equation}
where the summations on $i,j,k$ run from $0$ to $d^{(1)}_S - 1$, $d^{(2)}_S - 1$ and $d^{(3)}_S - 1$ respectively. Using the definition (\ref{4statoCS}) and the constraint (\ref{4constraint2}), for the relative state $\ket{\varphi(x^{(1)} , x^{(2)}, x^{(3)})}_{C,S} = \braket{x^{(1)} , x^{(2)}, x^{(3)}|\Psi}$ we can now obtain:

\begin{multline}
i\left(\frac{\partial}{\partial x^{(1)}} + \frac{\partial}{\partial x^{(2)}} + \frac{\partial}{\partial x^{(3)}} \right)\ket{\varphi(x^{(1)} , x^{(2)}, x^{(3)})}_{C,S}= \\
= \sum_{i,j,k} c_{ijk} p^{(1)}_i e^{ip^{(1)}_i x^{(1)}} e^{ip^{(2)}_j x^{(2)}} e^{ip^{(3)}_k x^{(3)}} \ket{-\epsilon_{ijk}}_C \otimes \ket{p^{(1)}_i , p^{(2)}_j, p^{(3)}_k }_S  + \\
+ \sum_{i,j,k} c_{ijk} p^{(2)}_j e^{ip^{(1)}_i x^{(1)}} e^{ip^{(2)}_j x^{(2)}} e^{ip^{(3)}_k x^{(3)}} \ket{-\epsilon_{ijk}}_C \otimes \ket{p^{(1)}_i , p^{(2)}_j, p^{(3)}_k }_S + \\
+ \sum_{i,j,k} c_{ijk} p^{(3)}_k e^{ip^{(1)}_i x^{(1)}} e^{ip^{(2)}_j x^{(2)}} e^{ip^{(3)}_k x^{(3)}} \ket{-\epsilon_{ijk}}_C \otimes \ket{p^{(1)}_i , p^{(2)}_j, p^{(3)}_k }_S = \\ = \left(\hat{P}^{(1)}_S + \hat{P}^{(2)}_S +\hat{P}^{(3)}_S \right) \ket{\varphi(x^{(1)} , x^{(2)}, x^{(3)})}_{C,S}
\end{multline}
which shows $\vec{P}_S=(\hat{P}^{(1)}_S,\hat{P}^{(2)}_S,\hat{P}^{(3)}_S)$ to be the generator of translations for the states of $C+S$ acting on the coordinates of the spatial reference frame $\vec{x} = (x^{(1)} , x^{(2)}, x^{(3)})$.

Expanding the state $\ket{\Psi}$ simultaneously on the coordinates basis $\left\{ \ket{x^{(1)} , x^{(2)}, x^{(3)} }_R\right\}$ and on the time basis $\left\{\ket{t}_C\right\}$ we find:
\begin{equation}\label{4esp}
\begin{split}
\ket{\Psi} &= A \left( \int dt \ket{t}\bra{t} \otimes   \int dx^{(1)} \int dx^{(2)} \int dx^{(3)} \ket{x^{(1)} , x^{(2)}, x^{(3)}} \bra{x^{(1)} , x^{(2)}, x^{(3)}} \right) \ket{\Psi} =\\&
= A \int dt  \int dx^{(1)} \int dx^{(2)} \int dx^{(3)} \ket{t}_C\otimes \ket{x^{(1)} , x^{(2)}, x^{(3)}}_R \otimes \ket{\psi(t,x^{(1)} , x^{(2)}, x^{(3)})}_S
\end{split}
\end{equation}
where $A= \frac{1}{T}\frac{1}{L^{(1)}_R} \frac{1}{L^{(2)}_R} \frac{1}{L^{(3)}_R}$. In equation (\ref{4esp}) the integral on time is evaluated from $t_0$ to $t_0 + T$ and each integral on $dx^{(J)}$ is evaluated from $x^{(J)}_0$ to $x^{(J)}_0 + L^{(J)}_R$. The state $\ket{\psi(t,x^{(1)} , x^{(2)}, x^{(3)})}_S = (\bra{t}\otimes\bra{x^{(1)} , x^{(2)}, x^{(3)}})\ket{\Psi}$ 
takes now the form 

\begin{equation}
\ket{\psi(t,x^{(1)} , x^{(2)}, x^{(3)})}_S = \sum_{i,j,k} c_{ijk}  e^{-it\epsilon_{ijk}} e^{-ip^{(1)}_i x^{(1)}} e^{-ip^{(2)}_j x^{(2)}} e^{-ip^{(3)}_k x^{(3)}} \ket{p^{(1)}_i , p^{(2)}_j, p^{(3)}_k }_S 
\end{equation}
where the summations on $i,j,k$ run again from $0$ to $d^{(1)}_S - 1$, $d^{(2)}_S - 1$ and $d^{(3)}_S - 1$ respectively.
Through this state we can calculate the conditional probability density of having the position $(y^{(1)},y^{(2)},y^{(3)})$ on $S$ at time $t$ and knowing that the spatial reference frame $R$ is in $(x^{(1)} , x^{(2)}, x^{(3)})$. We have: 

\begin{multline}
P(y^{(1)},y^{(2)},y^{(3)}\:|\:x^{(1)} , x^{(2)}, x^{(3)},t)
=  	\frac{1}{L^{(1)}_S} \frac{1}{L^{(2)}_S} \frac{1}{L^{(3)}_S} \left| \braket{y^{(1)},y^{(2)},y^{(3)}|\psi(t,x^{(1)} , x^{(2)}, x^{(3)})} \right|^2 = \\
= 	\frac{1}{L^{(1)}_S} \frac{1}{L^{(2)}_S} \frac{1}{L^{(3)}_S} \left| \sum_{i,j,k}
c_{ijk} e^{-it \epsilon_{ijk}} e^{ip^{(1)}_i(y^{(1)}  -  x^{(1)} )} e^{ip^{(2)}_j (y^{(2)}  -  x^{(2)} )} e^{ip^{(3)}_k(y^{(3)}  -  x^{(3)} )} \right|^2 .
\end{multline}
As in the previous cases, this probability density is well-defined (indeed we have $\int dy^{(1)} \int dy^{(2)} \int dy^{(3)} P(y^{(1)},y^{(2)},y^{(3)}\:|\:x^{(1)} , x^{(2)}, x^{(3)},t)=1$ with each integral on $dy^{(J)}$ evaluated from $y^{(J)}_0$ to $y^{(J)}_0 + L^{(J)}_S$) and depends on time and on the relative distance between $R$ and $S$ in a 3-dimensional continuous~space.

	\subsection{Free Particles (with $M\gg m$) in $3+1$ Spacetime}
We give here a simple example of an emerging $3+1$ dimensional spacetime using continuous values of space and time, by starting from the framework described in Section V.B. In doing so we adopt a slightly different formalism, which allows us to emphasize how space and time are treated here on equal footing. 

We consider two systems that we call $\mathfrak{R}$ and $S$. The system $\mathfrak{R}$ acts as spacetime reference frame for $S$ and it is composed of a free particle of mass $M$ together with an additional degree of freedom (with zero momentum) that acts as a clock. We choose also $S$ as free particle of mass $m$ and we assume $M \gg m$ (as mentioned in Section III.B this choice implies that we will be able to neglect the kinetic energy term of $\mathfrak{R}$ with respect to the kinetic energy of $S$, namely $\mathfrak{R}$ is a good reference frame and moves very slightly in time). 
The global Hamiltonian can be written:

\begin{equation}
\hat{H}=\hat{H}^{(0)}_{\mathfrak{R}} +\hat{H}^{(1)}_{\mathfrak{R}}+ \hat{H}^{(2)}_{\mathfrak{R}} + \hat{H}^{(3)}_{\mathfrak{R}} +\hat{H}_{S}
\end{equation}
where $\hat{H}^{(0)}_{\mathfrak{R}}$ is the Hamiltonian of the temporal reference frame (which takes the place of what was $\hat{H}_C$ in the previous discussion), $\hat{H}^{(1)}_{\mathfrak{R}}$, $\hat{H}^{(2)}_{\mathfrak{R}}$, $\hat{H}^{(3)}_{\mathfrak{R}}$ are the Hamiltonians depending on the momenta of the reference frame in the three spatial directions through $\hat{H}^{(1)}_{\mathfrak{R}} = \left( \hat{P}^{(1)}_{\mathfrak{R}} \right)^2/2M$, $\hat{H}^{(2)}_{\mathfrak{R}} = \left( \hat{P}^{(2)}_{\mathfrak{R}} \right)^2/2M$, $\hat{H}^{(3)}_{\mathfrak{R}} = \left( \hat{P}^{(3)}_{\mathfrak{R}} \right)^2/2M$ and $\hat{H}_S = \hat{H}^{(1)}_S + \hat{H}^{(2)}_S+ \hat{H}^{(3)}_S$ is the Hamiltonian of $S$ depending on the momenta $\hat{P}^{(1)}_S$, $\hat{P}^{(2)}_S$, $\hat{P}^{(3)}_S$ through $\hat{H}^{(1)}_{S} = \left( \hat{P}^{(1)}_{S} \right)^2/2m$, $\hat{H}^{(2)}_{S} = \left( \hat{P}^{(2)}_{S} \right)^2/2m$, $\hat{H}^{(3)}_{S} = \left( \hat{P}^{(3)}_{S} \right)^2/2m$. 
With the intent of treating space and time on equal footing, within the time subspace we rewrite the time states as $\ket{x^{(0)}} = \sum_{k=0}^{d^{(0)}_{\mathfrak{R}} -1} e^{-ix^{(0)} p^{(0)}_k} \ket{p^{(0)}_k}$ and the time values as $x^{(0)} \in \left[ x^{(0)}_0, x^{(0)}_0 + L^{(0)}_{\mathfrak{R}} \right]$, where $d^{(0)}_{\mathfrak{R}}$ is the dimension of the time subspace, $p^{(0)}_k$ are the eigenvalues of $\hat{H}^{(0)}_{\mathfrak{R}}$ and where $L^{(0)}_{\mathfrak{R}}$ takes now the place of what was $T$ in the previous discussion. Furthermore we redefine the position states as $\ket{x^{(J)}} = \sum_{k=0}^{d^{(J)}_{\mathfrak{R}} -1} e^{i x^{(J)} p^{(J)}_k} \ket{p^{(J)}_k}$ in $\mathfrak{R}$ and $\ket{y^{(J)}} = \sum_{k=0}^{d^{(J)}_{S} -1} e^{i y^{(J)} p^{(J)}_k} \ket{p^{(J)}_k}$ in $S$ for $J=1,2,3$.     

The constraints (\ref{4constraint1}) and (\ref{4constraint2}) read now:

\begin{equation}\label{5constraint1}
\hat{H}\ket{\Psi} = \left( \hat{H}^{(0)}_{\mathfrak{R}} +\hat{H}^{(1)}_{\mathfrak{R}}+ \hat{H}^{(2)}_{\mathfrak{R}} + \hat{H}^{(3)}_{\mathfrak{R}} +\hat{H}_{S} \right)\ket{\Psi}=0
\end{equation}
and

\begin{equation}\label{5constraint2}
\begin{split}
\vec{P}\ket{\Psi}	= \left( \vec{P}_{\mathfrak{R}} + \vec{P}_S \right)\ket{\Psi} = 0
\end{split}
\end{equation} 
with $ (\hat{P}^{(1)}_{\mathfrak{R}} + \hat{P}^{(1)}_S)\ket{\Psi}=(\hat{P}^{(2)}_{\mathfrak{R}} + \hat{P}^{(2)}_S)\ket{\Psi}= (\hat{P}^{(3)}_{\mathfrak{R}} + \hat{P}^{(3)}_S)\ket{\Psi}=0$. Assuming again $d^{(0)}_{\mathfrak{R}}, d^{(1)}_{\mathfrak{R}},d^{(2)}_{\mathfrak{R}},d^{(3)}_{\mathfrak{R}} \gg d^{(1)}_S,d^{(2)}_S,d^{(3)}_S$, the global state $\ket{\Psi}$ simultaneously satisfying (\ref{5constraint1}) and (\ref{5constraint2}) can be written as

\begin{equation}
\ket{\Psi} = \sum_{i=0}^{d^{(1)}_S - 1} \sum_{j=0}^{d^{(2)}_S - 1} \sum_{k=0}^{d^{(3)}_S - 1} c_{ijk} \ket{p^{(0)}=- \epsilon_{ijk}, -p^{(1)}_i , -p^{(2)}_j, -p^{(3)}_k}_{\mathfrak{R}} \otimes \ket{p^{(1)}_i , p^{(2)}_j, p^{(3)}_k }_S
\end{equation}
where $p^{(0)}$ is the value for the energy of the time reference and the energy function $\epsilon_{ijk}$ is: $\epsilon_{ijk} = \left(\frac{1}{2M} + \frac{1}{2m}\right) \left( \left(p^{(1)}_i\right)^{2} + \left(p^{(2)}_j\right)^{2} + \left(p^{(3)}_k\right)^{2}\right) \simeq \frac{1}{2m}\left( \left(p^{(1)}_i\right)^{2} + \left(p^{(2)}_j\right)^{2} + \left(p^{(3)}_k\right)^{2}\right)$. 

We can now expand the global state $\ket{\Psi}$ in the basis $\left\{\ket{x^{(0)},x^{(1)},x^{(2)},x^{(3)}}_{\mathfrak{R}} \right\}$ in the space $\mathfrak{R}$, thus obtaining:

\begin{equation}
\begin{split}
\ket{\Psi} &= A \int dx^{(0)} \int dx^{(1)} \int dx^{(2)} \int dx^{(3)} \ket{x^{(0)},x^{(1)},x^{(2)},x^{(3)}}\braket{x^{(0)},x^{(1)},x^{(2)},x^{(3)}|\Psi} =\\&
= A\int dx^{(0)} \int dx^{(1)} \int dx^{(2)} \int dx^{(3)} \ket{x^{(0)},x^{(1)},x^{(2)},x^{(3)}}_{\mathfrak{R}} \otimes \ket{\psi(x^{(0)},x^{(1)},x^{(2)},x^{(3)})}_S
\end{split}
\end{equation}
where $A= \frac{1}{L^{(0)}_{\mathfrak{R}}} \frac{1}{L^{(1)}_{\mathfrak{R}}} \frac{1}{L^{(2)}_{\mathfrak{R}}} \frac{1}{L^{(3)}_{\mathfrak{R}}}$ and the integrals on $dx^{(J)}$ are evaluated from $x^{(J)}_0$ to $x^{(J)}_0 + L^{(J)}_{\mathfrak{R}}$ for $J=0,1,2,3$. The state $ \ket{\psi(x^{(0)},x^{(1)},x^{(2)},x^{(3)})}_S = \braket{x^{(0)},x^{(1)},x^{(2)},x^{(3)}|\Psi}$ is the relative state of the system $S$ conditioned on the value $(x^{(0)},x^{(1)},x^{(2)},x^{(3)})$ for the spacetime reference frame $\mathfrak{R}$ and it takes the form:

\begin{multline}\label{5statoS}
\ket{\psi(x^{(0)},x^{(1)} , x^{(2)}, x^{(3)})}_S \simeq \sum_{i=0}^{d^{(1)}_S - 1} \sum_{j=0}^{d^{(2)}_S -1} \sum_{k=0}^{d^{(3)}_S - 1} c_{ijk}  e^{-i \frac{1}{2m}\left( (p^{(1)}_i)^{2} + (p^{(2)}_j)^{2} + (p^{(3)}_k)^{2}\right) x^{(0)}} \\ \times e^{ip^{(1)}_i x^{(1)}} e^{ip^{(2)}_j x^{(2)}} e^{ip^{(3)}_k x^{(3)}} \ket{p^{(1)}_i , p^{(2)}_j, p^{(3)}_k }_S .
\end{multline}
For the relative state (\ref{5statoS}), through (\ref{5constraint1}) and (\ref{5constraint2}), we easily find:
\begin{equation}\label{evfinale1}
i \frac{\partial}{\partial x^{(0)}} \ket{\psi(x^{(0)}, x^{(1)} , x^{(2)}, x^{(3)})}_S \simeq \hat{H}_S \ket{\psi(x^{(0)},x^{(1)} , x^{(2)}, x^{(3)})}_S
\end{equation}
and 
\begin{equation}\label{evfinale2}
- i\frac{\partial}{\partial x^{(J)}} \ket{\psi(x^{(0)},x^{(1)} , x^{(2)}, x^{(3)})}_{S} = \hat{P}^{(J)}_S \ket{\psi(x^{(0)},x^{(1)} , x^{(2)}, x^{(3)})}_{S} 
\end{equation}
for $J=1,2,3$. Equations (\ref{5statoS}), (\ref{evfinale1}) and (\ref{evfinale2}) lead to (writing $\vec{x}=(x^{(1)} , x^{(2)}, x^{(3)})$): 

\begin{equation}\label{evfinale3}
i \frac{\partial}{\partial x^{(0)}} \ket{\psi(x^{(0)}, \vec{x})}_S \simeq - \frac{1}{2m}\left(\frac{\partial^{2}}{\left(\partial x^{(1)}\right)^2} + \frac{\partial^{2}}{\left(\partial x^{(2)}\right)^2} + \frac{\partial^{2}}{\left(\partial x^{(3)}\right)^2} \right)\ket{\psi(x^{(0)},\vec{x})}_{S}
\end{equation}
which describes the dynamics of the particle in $S$ with respect to the coordinates of the $3+1$ dimensional quantum reference frame. 
We emphasize here that the formalism adopted allows us to write the equation (\ref{evfinale3}) for the state $\ket{\psi(x^{(0)},x^{(1)} , x^{(2)}, x^{(3)})}_S$ because the values of time and space of the subspace $\mathfrak{R}$ enter as parameters in $S$ thanks to the entanglement present in the global state $\ket{\Psi}$.

\subsection{System $S$ as a Relativistic Particle}
The formalism adopted in the previous paragraph is particularly well suited in describing the behavior of a relativistic particle. Considering indeed the system $S$ to be a relativistic particle with no internal degree of freedom (namely with spin $0$), we have for the energy function ($c=1$):

\begin{equation}\label{KGenergy}
\epsilon_{ijk} \simeq \pm \sqrt{(p^{(1)}_i)^2 + (p^{(2)}_j)^2 +(p^{(3)}_k)^2 + m^2} 
\end{equation}
which can be obtained from the energy constraint 
$\left( \left(\hat{H}^{(0)}_{\mathfrak{R}} \right)^2 - \left|\vec{P}_{S}\right|^2 -m^2 \right)\ket{\Psi}\simeq 0$
(where $\vec{P}_{S}=\left(\hat{P}^{(1)}_{S},\hat{P}^{(2)}_{S},\hat{P}^{(3)}_{S} \right)$ and the kinetic energy term of $\mathfrak{R}$ has been neglected).
The relative state (\ref{5statoS}) of the system $S$ conditioned on the value $(x^{(0)},x^{(1)},x^{(2)},x^{(3)})$ of the spacetime reference frame $\mathfrak{R}$ reads now:

\begin{multline}\label{5statoS2}
\ket{\psi_{\pm}(x^{(0)},x^{(1)} , x^{(2)}, x^{(3)})}_S \simeq \sum_{i=0}^{d^{(1)}_S - 1} \sum_{j=0}^{d^{(2)}_S -1} \sum_{k=0}^{d^{(3)}_S - 1} c_{ijk}  e^{\mp i x^{(0)} \sqrt{(p^{(1)}_i)^2 + (p^{(2)}_j)^2 +(p^{(3)}_k)^2 + m^2} } \\ \times e^{ip^{(1)}_i x^{(1)}} e^{ip^{(2)}_j x^{(2)}} e^{ip^{(3)}_k x^{(3)}} \ket{p^{(1)}_i , p^{(2)}_j, p^{(3)}_k }_S 
\end{multline}
where we have considered together both the results for the energy function (\ref{KGenergy}). 
For the relative state (\ref{5statoS2}) we find:

\begin{equation}\label{5.1}
\frac{\partial^2}{\partial (x^{(0)})^2}  \ket{\psi_{\pm}(x^{(0)},x^{(1)} , x^{(2)}, x^{(3)})}_S  \simeq - \left( \left(\hat{P}^{(1)}_S\right)^2 + \left(\hat{P}^{(2)}_S\right)^2 + \left(\hat{P}^{(3)}_S\right)^2 + m^2 \right) \ket{\psi_{\pm}(x^{(0)},x^{(1)} , x^{(2)}, x^{(3)})}_S
\end{equation}
and
\begin{equation}\label{5.2}
\begin{split}
\frac{\partial^2}{\partial (x^{(J)})^2} \ket{\psi_{\pm}(x^{(0)},x^{(1)} , x^{(2)}, x^{(3)})}_S = - \left(\hat{P}^{(J)}_S\right)^2 \ket{\psi_{\pm}(x^{(0)},x^{(1)} , x^{(2)}, x^{(3)})}_S
\end{split}
\end{equation}
with $J=1,2,3$. Through equations (\ref{5.1}) and (\ref{5.2}) we easily obtain: 

\begin{equation}\label{KG}
\left(\frac{\partial^2}{\partial (x^{(0)})^2} -  \frac{\partial^2}{\partial (x^{(1)})^2} -  \frac{\partial^2}{\partial (x^{(2)})^2} - \frac{\partial^2}{\partial (x^{(3)})^2} +m^2 \right) \ket{\psi_{\pm}(x^{(0)}, \vec{x})}_S \simeq 0
\end{equation}
which describes the dynamics of the particle in $S$ with respect to the coordinates of the $3+1$ dimensional quantum reference frame. Equation (\ref{KG}) has the form of the Klein-Gordon equation but differs from it being the derivatives applied to the state $\ket{\psi_{\pm}(x^{(0)},x^{(1)} , x^{(2)}, x^{(3)})}_S$ and not to the wave function. Also in this case, this is possible since the time and space values of $\mathfrak{R}$ enter as parameters in the state of the subsystem $S$ thanks to the entanglement present in the global state $\ket{\Psi}$.

A similar result can be obtained considering the system $S$ as a relativistic particle with spin $1/2$. In this case the global state of the Universe can be written:

\begin{equation}
\ket{\Psi} = \sum_{\sigma = 0}^{3} \sum_{n=0}^{d^{(0)}_{\mathfrak{R}} - 1} \sum_{i=0}^{d^{(1)}_S - 1} \sum_{j=0}^{d^{(2)}_S - 1} \sum_{k=0}^{d^{(3)}_S - 1} c_{nijk} \ket{p^{(0)}_n , -p^{(1)}_i , -p^{(2)}_j, -p^{(3)}_k}_{\mathfrak{R}} \otimes \ket{p^{(1)}_i , p^{(2)}_j, p^{(3)}_k,\sigma}_S
\end{equation} 
where we have introduced the spin degree of freedom within the subspace $S$ in accordance to Refs. \onlinecite{dirac,librodirac} and where the value of $p^{(0)}_n$ is constrained through 

\begin{equation}\label{dirac1}
\left( \hat{H}^{(0)}_{\mathfrak{R}} + \vec{\alpha} \cdot \vec{P}_S + \beta m \right)\ket{\Psi} \simeq 0 .
\end{equation}
In equation (\ref{dirac1}) we have written for the system $S$ the free Dirac Hamiltonian as $\hat{H}_S = \vec{\alpha}\cdot \vec{P}_S + \beta m$ \cite{dirac} and we have again neglected the kinetic energy term of $\mathfrak{R}$. The state $\ket{\psi(x^{(0)},x^{(1)} , x^{(2)}, x^{(3)})}_S = \braket{x^{(0)},x^{(1)} , x^{(2)}, x^{(3)}|\Psi}$ reads now:

\begin{multline}\label{diracrel}
\ket{\psi (x^{(0)},x^{(1)} , x^{(2)}, x^{(3)})}_S = \sum_{\sigma = 0}^{3} \sum_{n=0}^{d^{(0)}_{\mathfrak{R}} - 1} \sum_{i=0}^{d^{(1)}_S - 1} \sum_{j=0}^{d^{(2)}_S -1} \sum_{k=0}^{d^{(3)}_S - 1} c_{nijk} e^{- ip^{(0)}_n x^{(0)}} \\ \times e^{ip^{(1)}_i x^{(1)}} e^{ip^{(2)}_j x^{(2)}} e^{ip^{(3)}_k x^{(3)}} \ket{p^{(1)}_i , p^{(2)}_j, p^{(3)}_k,\sigma}_S.
\end{multline}
For the relative state (\ref{diracrel}) still holds

\begin{equation}\label{dirac2}
i \frac{\partial}{\partial x^{(0)}} \ket{\psi(x^{(0)},x^{(1)} , x^{(2)}, x^{(3)})}_S \simeq \hat{H}_S \ket{\psi(x^{(0)},x^{(1)} , x^{(2)}, x^{(3)})}_S
\end{equation}
and
\begin{equation}\label{dirac3}
- i \frac{\partial}{\partial x^{(J)}} \ket{\psi(x^{(0)},x^{(1)} , x^{(2)}, x^{(3)})}_{S} = \hat{P}^{(J)}_S   \ket{\psi(x^{(0)},x^{(1)} , x^{(2)}, x^{(3)})}_{S} .
\end{equation}
So, starting from equations (\ref{dirac2}) and (\ref{dirac3}), writing $\vec{\alpha}=(\alpha^{(1)}, \alpha^{(2)}, \alpha^{(3)})$ and remembering that $\hat{H}_S = \vec{\alpha}\cdot \vec{P}_S + \beta m$, we obtain:

\begin{equation}
i \frac{\partial}{\partial x^{(0)}} \ket{\psi(x^{(0)},\vec{x})}_S \simeq  \left(-i \alpha^{(1)}\frac{\partial}{\partial x^{(1)}} -i \alpha^{(2)}\frac{\partial}{\partial x^{(2)}} -i \alpha^{(3)}\frac{\partial}{\partial x^{(3)}} + \beta m   \right)  \ket{\psi(x^{(0)},\vec{x})}_S
\end{equation}
\end{widetext} 
which has the form of the Dirac equation and again describes the dynamics of the particle in $S$ with respect to the coordinates of the $3+1$ dimensional quantum reference frame. All the considerations made for equation (\ref{KG}) still apply in this case. Clearly, in order to give a complete relativistic generalization of the model, in addition to this discussion, we need to consider relativistic reference frames and a protocol that allows to change the point of view between different observers in different reference frames (so that dilation of times and contraction of lengths can be derived), but this is beyond the scope of the present work.

	\section{Conclusions}
\label{Conclusions}
The PaW mechanism was originally introduced in order to describe the emergence of time from entanglement.
In this work we first extended the PaW mechanism at the spatial degree of freedom and then we provide a description of a model of non-relativistic quantum spacetime. In doing this we started focusing on space and we showed that, in a closed quantum system satisfying a global constraint on total momentum (and therefore with the absolute position totally indeterminate), a well-defined relative position emerge between the subsystems $S$ and $R$, where the latter is taken as quantum spatial reference frame. In the spaces $R$ and $S$, generalizing the approach outlined in Refs. \onlinecite{nostro,nostro2,pegg}, we considered non-degenerate, discrete spectra for the momentum operators and we introduce POVMs in describing the spatial degrees of freedom. In this way we recovered continuous values of space in $S$ and $R$ also for a discrete momentum spectra (the case of momentum with continuous spectrum was then also treated in Section III.E). Finally we introduced in the Universe an additional subsystem $C$ acting as a clock and we considered the Universe satisfying a double constraint: both on total momentum and total energy. We showed how this framework can be implemented without contradiction in the simple case of one spatial degree of freedom (considering also the case of multiple time measurements) and in the \lq\lq more physical\rq\rq case of three spatial degrees of freedom thus providing a $3+1$ dimensional quantum spacetime emerging from entanglement.


\begin{acknowledgments}
We acknowledge funding from the H2020 QuantERA ERA-NET Cofund in Quantum Technologies projects QCLOCKS. 
\end{acknowledgments}




\appendix

\begin{widetext}

\section{Proof of Equation (\ref{conditionalprobabilitydiscreta})}
We start considering the global state written as
\begin{equation}
\ket{\Psi} =  \frac{\sqrt{d_R}}{D_R} \sum_{j=0}^{D_R-1} \ket{x_j}_R\otimes\ket{\phi(x_j)}_S 
\end{equation}
where $\ket{\phi(x_j)}_S = \sum_{k=0}^{d_S-1} c_k e^{-i p_k x_j} \ket{p_k}_S$. We can now calculate the conditional probability as follows
\begin{equation}
\begin{split}
& P(y_l \: on \: S \:|\: x_j \: on \: R) =\frac{d_S}{D_S} \frac{\braket{\Psi|x_j}\bra{x_j}\otimes\ket{y_l}\braket{y_l|\Psi}}{\braket{\Psi|x_j}\braket{x_j|\Psi}} =\\ \\&
= \frac{d_S}{D_S} \frac{\frac{d_R}{D^{2}_R} \sum_{n}\sum_{m} \braket{x_n|x_j}\braket{x_j|x_m} \braket{\phi(x_n)|y_l} \braket{y_l|\phi(x_m)}}{\frac{d_R}{D^{2}_R} \sum_{n'}\sum_{m'} \braket{x_{n'}|x_j}\braket{x_j|x_{m'}} \braket{\phi(x_{n'})|\phi(x_{m'})}}=\\ \\&
= \frac{d_S}{D_S} \frac{\sum_{n}\sum_{m} \braket{x_n|x_j}\braket{x_j|x_m} \sum_{k}\sum_{a}c^{*}_kc_a e^{ip_k x_n} e^{-i p_a x_m} \braket{p_k|y_l} \braket{y_l|p_a}}{\sum_{n'}\sum_{m'} \braket{x_{n'}|x_j}\braket{x_j|x_{m'}} \sum_{b}\sum_{k'}c^{*}_b c_{k'} e^{ip_b x_{n'}} e^{-i p_{k'}x_{m'}} \braket{p_b|p_{k'}}    }= \\ \\&
=\frac{1}{D_S} \frac{\sum_{n}\sum_{m}  \sum_{i}\sum_{g}e^{- p_i(x_j - x_n)} e^{ip_g(x_j - x_m)}      \sum_{k}\sum_{a}c^{*}_kc_a e^{ip_k x_n} e^{-i p_a x_m} e^{-i y_l p_k} e^{i y_l p_a}}{\sum_{n'}\sum_{m'}    \sum_{i'}\sum_{g'}e^{- p_{i'}(x_j - x_{n'})} e^{ip_{g'}(x_j - x_{m'})}      \sum_{k'} |c_{k'}|^2 e^{ip_{k'} (x_{n'} -x_{m'})}     }
\end{split}
\end{equation}

We use now the fact that (see Appendix C for the proof)
\begin{equation}\label{delta2}
\sum_{j=0}^{D_R -1} e^{-i x_j(p_k - p_n)} = D_R\delta_{p_k,p_n}
\end{equation}
and so we obtain
\begin{equation}
P(y_l \: on \: S \:|\: x_j \: on \: R) = \frac{1}{D_S} \frac{\sum_{k}\sum_{a} c^{*}_k c_a e^{i(p_a - p_k)(y_l-x_j)}}{\sum_{k'} |c_{k'}|^2} =  \frac{d_S}{D_S}\left| \braket{y_l|\phi(x_j)} \right|^2 = \frac{1}{D_S} \left| \sum_{k=0}^{d_S-1} c_k e^{i p_k(y_l-x_j)} \right|^2
\end{equation}
where we have considered that $\sum_{k'} |c_{k'}|^2 = 1$.

\section{Proof of Equation (\ref{conditionalprobability})}
We start considering the global state written as
\begin{equation}
\ket{\Psi} = \frac{1}{L_R} \int_{x_0}^{x_0 + L_R} d x   \ket{x}_R \otimes \ket{\phi(x)}_S 
\end{equation}
where $\ket{\phi(x)}_S = \sum_{k=0}^{d_S-1} c_k e^{-i p_k x} \ket{p_k}_S$. We can now calculate the conditional probability density as follows (all the integrals are evaluated between $x_0$ and $x_0+L_R$):
\begin{equation}
\begin{split}
& P(y \: on \: S \:|\: x \: on \: R)  = \frac{1}{L_S} \frac{\braket{\Psi|x}\bra{x}\otimes\ket{y}\braket{y|\Psi}}{\braket{\Psi|x}\braket{x|\Psi}} = \\ \\&
= \frac{1}{L_S} \frac{\frac{1}{L^{2}_R} \int dx' \int dx^{''} \braket{x'|x}\braket{x|x^{''}} \braket{\phi(x')|y}\braket{y|\phi(x^{''})}}{\frac{1}{L^{2}_R} \int dx' \int dx^{''} \braket{x'|x}\braket{x|x^{''}} \braket{\phi(x')|\phi(x^{''})}} = \\ \\&
= \frac{1}{L_S} \frac{  \int dx' \int dx^{''} \sum_{m}\sum_{l}e^{-ip_m(x-x')}e^{ip_l(x-x^{''})} \sum_{k}\sum_{n} c^{*}_k c_n e^{ip_k x' }e^{-ip_n x^{''}} \braket{p_k|y}\braket{y|p_n}}{ \int dx' \int dx^{''} \sum_{m'}\sum_{l'}e^{-ip_{m'} (x-x')}e^{ip_{l'}(x-x^{''})} \sum_{k'}\sum_{n'} c^{*}_{k'} c_{n'} e^{ip_{k'} x' }e^{-ip_{n'} x^{''}}\braket{p_{k'}|p_{n'}}} = \\ \\&
= \frac{1}{L_S} \frac{  \int dx' \int dx^{''} \sum_{m}\sum_{l}e^{-ip_m(x-x')}e^{ip_l(x-x^{''})} \sum_{k}\sum_{n} c^{*}_k c_n e^{ip_k x' }e^{-ip_n x^{''}}  e^{-ip_k y} e^{ip_n y}          }{   \int dx' \int dx^{''} \sum_{m'}\sum_{l'}e^{-ip_{m'} (x-x')}e^{ip_{l'}(x-x^{''})} \sum_{k'} |c_{k'}|^2 e^{ip_{k'} (x' - x^{''})}} .
\end{split}
\end{equation}

We use now the fact that (see Appendix C for the proof)
\begin{equation}
\int_{x_0}^{x_0 + L_R} d x' e^{-i x' (p_k - p_n)} = L_R \delta_{p_k,p_n} ,
\end{equation}
thus we obtain:
\begin{equation}
\begin{split}
P(y \: on \: S \:|\: x \: on \: R) = \frac{1}{L_S} \frac{\sum_{k}\sum_{n} c^{*}_k c_n e^{i(p_n - p_k)(y-x)}}{\sum_{k'} |c_{k'}|^2}  = \frac{1}{L_S}\left| \braket{y|\phi(x)} \right|^2 = \frac{1}{L_S} \left| \sum_{k=0}^{d_S-1} c_k e^{i p_k(y-x)} \right|^2
\end{split}
\end{equation}
where again we have considered that $\sum_{k'} |c_{k'}|^2 = 1$. 

\section{Proof of Equations (\ref{delta}) and (\ref{delta2})}
We first approach the case with discrete space values (namely we first demonstrate equation (\ref{delta2})). We start considering the global state $\ket{\Psi}$ written as $\ket{\Psi} = \sum_{k=0}^{d_S -1} c_k \ket{p=-p_k}_R\otimes\ket{p_k}_S$ 
and we apply in sequence the resolutions of the identity $\mathbb{1}_{R} = \frac{d_R}{D_R} \sum_{j=0}^{D_R -1} \ket{x_j}\bra{x_j}$ and $\mathbb{1}_{R} = \sum_{n=0}^{d_R -1} \ket{p_n}\bra{p_n}$.
We obtain
\begin{equation}
\begin{split}
\ket{\Psi} & =	\frac{d_R}{D_R} \sum_{j=0}^{D_R -1} \ket{x_{j}}\braket{x_{j}|\Psi}=  \\&
=	\frac{\sqrt{d_R}}{D_R} \sum_{j=0}^{D_R -1} \ket{x_{j}}_R\otimes \sum_{k=0}^{d_S -1} c_ke^{-ip_k x_{j}}\ket{p_k}_S = \\&
= \sum_{n=0}^{d_R -1} \ket{p_n}\bra{p_n}\frac{\sqrt{d_R}}{D_R} \sum_{j=0}^{D_R -1} \ket{x_{j}}_R\otimes \sum_{k=0}^{d_S -1} c_ke^{-ip_k x_{j}}\ket{p_k}_S =  \\&
= \sum_{n=0}^{d_R -1}\sum_{k=0}^{d_S -1}c_k \frac{1}{D_R} \sum_{j=0}^{D_R -1}   e^{-ix_j (p_n + p_k)} \ket{p_n}_R \otimes \ket{p_k}_S
\end{split}
\end{equation}
from which we have
\begin{equation}
\sum_{j=0}^{D_R -1} e^{-i x_j(p_k + p_n)} = D_R\delta_{p_n,-p_k} .
\end{equation}

Let us now consider the limiting case $z \longrightarrow \infty$, i.e. we assume the continuous representation of the coordinate $x$ which can now take any real value from $x_0$ to $x_0 + L_R$. We start again from the global state written as (\ref{miserveperlafine}) and we apply in sequence the resolutions of the identity
\begin{equation}
\mathbb{1}_{R} = \frac{1}{L_R} \int_{x_0}^{x_0+L_R} dx \ket{x} \bra{x}  \:\:\: and \:\:\: \mathbb{1}_{R} = \sum_{n=0}^{d_R -1} \ket{p_n}\bra{p_n} .
\end{equation}
We have:
\begin{equation}
\begin{split}
\ket{\Psi} & =	\frac{1}{L_R} \int_{x_0}^{x_0 + L_R} dx \ket{x}\braket{x|\Psi}=  \\&
=\frac{1}{L_R} \int_{x_0}^{x_0 + L_R} dx \ket{x}_R\otimes \sum_{k=0}^{d_S -1} c_ke^{-ip_kx}\ket{p_k}_S =  \\&
= \sum_{n=0}^{d_R -1} \ket{p_n}\bra{p_n} \frac{1}{L_R} \int_{x_0}^{x_0 + L_R} dx \ket{x}_R\otimes \sum_{k=0}^{d_S -1} c_k e^{-ip_kx}\ket{p_k}_S=  \\&
= \sum_{n=0}^{d_R -1}\sum_{k=0}^{d_S -1}c_k \frac{1}{L_R} \int_{x_0}^{x_0 + L_R} dx e^{-ix(p_n + p_k)} \ket{p_n}_R \otimes \ket{p_k}_S
\end{split}
\end{equation}
from which we can easily see that $\int_{x_0}^{x_0 + L_R} dx e^{- i x(p_n + p_k)} = L_R \delta_{p_n, -p_k}$.
\end{widetext}

\section{Emergent Spacetime with $\hat{P}_C \ne 0$}
We discuss here our model of $1+1$ dimensional spacetime in the case of $\hat{P}_C \ne 0$. The contraints for the global state $\ket{\Psi} \in \mathcal{H}_C\otimes\mathcal{H}_R\otimes\mathcal{H}_S$ are now:
\begin{equation}\label{1A}
\hat{H}\ket{\Psi} = (\hat{H}_C + \hat{H}_R + \hat{H}_S )\ket{\Psi}=0
\end{equation}
and
\begin{equation}\label{2A}
\hat{P}\ket{\Psi} = (\hat{P}_C + \hat{P}_R + \hat{P}_S )\ket{\Psi}=0 .
\end{equation}
As we mentioned in the main text, in this case, there could be limitations in the allowed momenta to ensure that (\ref{1A}) and (\ref{2A}) are together satisfied and the global state $\ket{\Psi}$ can not be written here in the simple form (\ref{statoglobalespaziotempo}). The limitations could arise from the fact that the energies and the momenta of $C$, $R$ and $S$ simultaneously have to sum to zero and the difficulty of the problem depends on the type of dispersion relations in the three subspaces. 
However, assuming $\ket{\Psi}$ satisfying the constraints, the discussion follows as in the case of $\hat{P}_C = 0$ that we have seen in Section III.B.

We expand also here the global state on the basis $\left\{\ket{t_m}_C\right\}$ in $C$ thanks to (\ref{pomidentity2}), thus obtaining
\begin{equation}\label{serveperGLMA}
\begin{split}
\ket{\Psi} &= \frac{d_C}{D_C} \sum_{m=0}^{D_C -1} \ket{t_m} \braket{t_m|\Psi} =\\&=  \frac{\sqrt{d_C}}{D_C} \sum_{m=0}^{D_C -1} \ket{t_m}_C \otimes \ket{\phi(t_m)}_{R,S}
\end{split}
\end{equation}
where $\ket{\phi(t_m)}_{R,S} = \sqrt{d_C} \braket{t_m|\Psi}$ is state of the composite system $R+S$ at time $t_m$. For such a state, through (\ref{1A}) and the relative state definition, it is easy to find again the time evolution with respect to the clock $C$:
\begin{equation}\label{evoluzioneRSA}
\begin{split}
\ket{\phi(t_m)}_{R,S} 
= e^{-i (\hat{H}_R + \hat{H}_S)(t_m - t_0)}\ket{\phi(t_0)}_{R,S}
\end{split}
\end{equation} 
where $\ket{\phi(t_0)}_{R,S}= \sqrt{d_C} \braket{t_0|\Psi}$ is the state of $R+S$ conditioned on $t_0$ that is the value of the clock taken as initial time. Equation (\ref{evoluzioneRSA}) shows, as expected, the simultaneous evolution of $R$ and $S$ over time. In the limiting case $s \longrightarrow \infty$ where $t$ takes all the real values between $t_0$ and $t_0+T$ the global state can again be written 
\begin{multline}
\ket{\Psi} = \frac{1}{T} \int_{t_0}^{t_0 +T} dt \ket{t} \braket{t|\Psi} =\\= \frac{1}{T} \int_{t_0}^{t_0 +T} dt \ket{t}_C \otimes \ket{\phi(t)}_{R,S}
\end{multline}
and defining the relative state of $R+S$ as $\ket{\phi(t)}_{R,S} = \braket{t|\Psi}$ we obtain:
\begin{equation}\label{evoluzioneRScA}
i \frac{\partial}{\partial t}\ket{\phi(t)}_{R,S} = \left(\hat{H}_R + \hat{H}_S\right)\ket{\phi(t)}_{R,S}
\end{equation}
that is the Schrödinger evolution for $R+S$ with respect to the clock time $t$, written in the differential form.

We can therefore expand the state $\ket{\Psi}$ in the coordinates $\left\{\ket{x_j}_R\right\}$ in $R$, thus obtaining:
\begin{equation}
\begin{split}
\ket{\Psi} &= \frac{d_R}{D_R} \sum_{j=0}^{D_R -1} \ket{x_j} \braket{x_j|\Psi} =\\&=  \frac{\sqrt{d_R}}{D_R} \sum_{j=0}^{D_R -1} \ket{x_j}_R \otimes \ket{\varphi(x_j)}_{C,S}
\end{split}
\end{equation}
where $\ket{\varphi(x_j)}_{C,S}=\sqrt{d_R}\braket{x_j|\Psi}$ is the relative state of $C+S$ conditioned to the value $x_j$ on the reference frame $R$. For the state $\ket{\varphi(x_j)}_{C,S}$ we find now:
\begin{equation}\label{mA}
	\begin{split}
\ket{\varphi(x_j)}_{C,S} & = \sqrt{d_R}\braket{x_j|\Psi} =\\&= \sqrt{d_R} \bra{x_0}e^{i\hat{P}_R (x_j -x_0)}\ket{\Psi} =  \\&
=        \sqrt{d_R} \bra{x_0}e^{i (\hat{P} - \hat{P}_C - \hat{P}_S )(x_j -x_0)}\ket{\Psi}  =\\&  = e^{- i ( \hat{P}_C + \hat{P}_S) (x_j -x_0)}\ket{\phi(x_0)}_{C,S}
\end{split}
\end{equation}
where the momentum of the clock $C$ appear in the equation since we have $\hat{P}_C \ne 0$. Also here we consider the limit $z \longrightarrow \infty$, where again $x$ can take all the real values between $x_0$ and $x_0 + L_R$. In this case the global state can be written
\begin{multline}\label{mmA}
\ket{\Psi} = \frac{1}{L_R} \int_{x_0}^{x_0 +L_R} dx \ket{x} \braket{x|\Psi} =\\= \frac{1}{L_R} \int_{x_0}^{x_0 +L_R} dx \ket{x}_R \otimes \ket{\varphi(x)}_{C,S}
\end{multline}
and, defining the relative state of $C+S$ as $\ket{\varphi(x)}_{C,S} = \braket{x|\Psi}$, we obtain 
\begin{equation}
\begin{split}
\left(\hat{P}_C + \hat{P}_S\right) \ket{\varphi(x)}_{C,S} &= \bra{x}\left( \hat{P}_C + \hat{P}_S \right)\ket{\Psi} =
\\&= \bra{x}\left( \hat{P} - \hat{P}_R \right)\ket{\Psi} =
\\&= - \sum_{k=0}^{d_R - 1} e^{i p_k x} \bra{p_k} \hat{P}_R \ket{\Psi} =
\\&= -  \left( \sum_{k=0}^{d_R - 1} p_k e^{i p_k x} \bra{p_k} \right) \ket{\Psi} =
\\&= i \frac{\partial}{\partial x} \ket{\varphi(x)}_{C,S} .
\end{split}
\end{equation} 
Through this latter equation and (\ref{mA}) we can see again that now the generator of translations in the coordinate values $x$ for the state $\ket{\varphi(x)}_{C,S}$ is the operator $\hat{P}_C + \hat{P}_S$.

Finally we can expand the state $\ket{\Psi}$ simultaneously on the coordinates $\left\{\ket{x_j}_R\right\}$ in $R$ and on the time basis $\left\{\ket{t_m}_C\right\}$ in $C$. We have for the global state:
\begin{equation}\label{45A}
\begin{split}
\ket{\Psi} &= \left(\frac{d_C}{D_C} \sum_{m=0}^{D_C -1} \ket{t_m}\bra{t_m} \otimes  \frac{d_R}{D_R} \sum_{j=0}^{D_R -1} \ket{x_j}\bra{x_j} \right)\ket{\Psi}=\\&
= \frac{\sqrt{d_C}}{D_C} \frac{\sqrt{d_R}}{D_R} \sum_{m=0}^{D_C -1}\sum_{j=0}^{D_R -1}\ket{t_m}_C\otimes\ket{x_j}_R\otimes\ket{\psi(t_m,x_j)}_S
\end{split}
\end{equation}
where $\ket{\psi(t_m,x_j)}_S =\sqrt{d_C} \sqrt{d_R}(\bra{t_m}\otimes\bra{x_j})\ket{\Psi}$ is the relative state of the system $S$ at time $t_m$ conditioned on the value $x_j$ for the reference frame $R$. Also here we can search the conditional probability of having a certain position $y_l$ in $S$ at time $t_m$ and knowing that the reference frame is in $x_j$, that is:
\begin{equation}\label{probfinalediscretaA}
P(y_l \: on\: S\:|\:x_j\:on\:R, \: t_m \: on \:C) = \frac{d_S}{D_S} |\braket{y_l|\psi(t_m,x_j)}|^2 
\end{equation}
that will depend on time $t_m$ and on the values $x_j$ and $y_l$ of $R$ and $S$ respectively.
Clearly we can extend these results also to the limiting cases $z,s \longrightarrow \infty$. Indeed we can write the global state $\ket{\Psi}$ as
\begin{equation}\label{47A}
\begin{split}
\ket{\Psi} &= \left( \frac{1}{T}\int_{t_0}^{t_0 + T} dt \ket{t}\bra{t} \otimes \frac{1}{L_R}\int_{x_0}^{x_0 + L_R} dx \ket{x}\bra{x} \right) \ket{\Psi} = \\&
= \frac{1}{T} \frac{1}{L_R} \int_{t_0}^{t_0 + T} dt \int_{x_0}^{x_0 + L_R} dx \ket{t}_C \otimes \ket{x}_R \otimes \ket{\psi(t,x)}_S
\end{split}
\end{equation}
where again $\ket{\psi(t,x)}_S = (\bra{t}\otimes\bra{x})\ket{\Psi}$ is the relative state of the system $S$ at time $t$ conditioned on the value $x$ for the reference frame $R$. The conditional probability density of having a certain position $y$ in $S$ at time $t$ and knowing that the reference frame is in $x$ is:
\begin{equation}\label{probfinaleA}
P(y \: on\: S\:|\:x\:on\:R,\: t \: on \:C) = \frac{1}{L_S}\left| \braket{y|\psi(t,x)} \right|^2  
\end{equation}
which will depend on time $t$ as well as on the position values of $x$ and $y$. We can not explicitly calculate the probabilities (\ref{probfinalediscretaA}) and (\ref{probfinaleA}) here since we can not write the state $\ket{\Psi}$ explicitly in the simple form (\ref{statoglobalespaziotempo}). 

In summary we assume as a \lq\lq good clock\rq\rq a system with $\hat{P}_C = 0$, a framework that can be easily implemented using an internal degree of freedom in describing the clock. If we want to use $\hat{P}_C \ne 0$ we have to find a state $\ket{\Psi}$ that satisfies (\ref{1A}) and (\ref{2A}), undergoing the limitations that this choice imposes and the discussion then follows as shown in this Appendix. A different (less elegant) framework might be to take $(\hat{P}_R + \hat{P}_S)\ket{\Psi} = 0$ as a constraint for the theory even in the case of a clock with non-zero momentum, but we believe that this choice (even if functional) does not have a good physical justification. In this latter case indeed the Universe would not be in an eigenstate of the global momentum and thus the symmetry with respect to the temporal degree of freedom (derived from a Universe in an eigenstate of the global Hamiltonian) would be lost.

\begin{widetext}
\section{Proof of Equation (\ref{probfinalediscreta})}
We start here considering the global state $\ket{\Psi}$ written as in equation \ref{statoglobalespaziotempo}, that is 
\begin{equation}
	\ket{\Psi} = \sum_{k=0}^{d_S -1} c_k \ket{E=-\epsilon_k}_C\otimes\ket{p=-p_k}_R\otimes\ket{p_k}_S
\end{equation}
where $\epsilon_k= E^{(R)}(-p_k) + E^{(S)}(p_k)$ is the energy function related to the momenta $p_k$ of $R$ and $S$. Now we write the conditional probability as:
\begin{equation}\label{75}
\begin{split}
& P(y_l \: on\: S\:|\:x_j\:on\:R,\: t_m \: on \:C) = \frac{d_S}{D_S} \frac{\braket{\Psi|t_m}\bra{t_m}\otimes\ket{x_j}\bra{x_j}\otimes\ket{y_l}\braket{y_l|\Psi}}{\braket{\Psi|t_m}\bra{t_m}\otimes\ket{x_j}\braket{x_j|\Psi}} =\\ \\&
= \frac{d_S}{D_S} \frac{\sum_{k}\sum_{n} c_k c^{*}_n \braket{E=-\epsilon_n|t_m}\braket{t_m|E=-\epsilon_k}\braket{p=-p_n|x_j}\braket{x_j|p=-p_k}\braket{p_n|y_l}\braket{y_l|p_k}}{\sum_{k'}\sum_{n'}c_{k'} c^{*}_{n'} \braket{E=-\epsilon_{n'}|t_m}\braket{t_m|E=-\epsilon_{k'}}\braket{p=-p_{n'}|x_j}\braket{x_j|p=-p_{k'}}\braket{p_{n'}|p_{k'}}} =\\ \\&
=\frac{1}{D_S}  \frac{\sum_{k}\sum_{n} c_k c^{*}_n e^{-it_m(\epsilon_k -\epsilon_n)} e^{i (p_k - p_n)(y_l-x_j)}}{\sum_{k'} |c_{k'}|^2} = \frac{1}{D_S} \left| \sum_{k=0}^{d_S -1} c_k e^{-i\epsilon_k t_m}e^{ip_k(y_l-x_j)} \right|^2
\end{split}
\end{equation}
where, in the last step, we have considered $\sum_{k'=0}^{d_S-1}|c_{k'}|^2 = 1$. From equation (\ref{75}) and considering $\ket{\psi(t_m,x_j)}_S= \sum_{k=0}^{d_S -1}c_k e^{-i\epsilon_k t_m} e^{-ip_k x_j}\ket{p_k}$, we can also see that the probability $P(y_l \: on\: S\:|\:x_j\:on\:R,\: t_m \: on \:C)$ can be written as $\frac{d_S}{D_S}|\braket{y_l|\psi(t_m,x_j)}|^2$.

\section{Proof of Equation (\ref{probfinale})}
We start also here considering the global state written as in \ref{statoglobalespaziotempo}, that is
\begin{equation}
\ket{\Psi} = \sum_{k=0}^{d_S -1} c_k \ket{E=-\epsilon_k}_C\otimes\ket{p=-p_k}_R\otimes\ket{p_k}_S
\end{equation}
where $\epsilon_k= E^{(R)}(- p_k) + E^{(S)}(p_k)$ is the energy function related to the momenta $p_k$ of $R$ and $S$. Now we write the probability density as:
\begin{equation}\label{76}
\begin{split}
& P(y \: on\: S\:|\:x\:on\:R,\: t \: on \:C) = \frac{1}{L_S} \frac{\braket{\Psi|t}\bra{t}\otimes\ket{x}\bra{x}\otimes\ket{y}\braket{y|\Psi}}{\braket{\Psi|t}\bra{t}\otimes\ket{x}\braket{x|\Psi}} =\\ \\&
=  \frac{1}{L_S} \frac{\sum_{k}\sum_{n} c_k c^{*}_n \braket{E=-\epsilon_n|t}\braket{t|E=-\epsilon_k}\braket{p=-p_n|x}\braket{x|p=-p_k}\braket{p_n|y}\braket{y|p_k}}{\sum_{m}\sum_{l}c_m c^{*}_l \braket{E=-\epsilon_l|t}\braket{t|E=-\epsilon_m}\braket{p=-p_l|x}\braket{x|p=-p_m}\braket{p_l|p_m}} =\\ \\&
= \frac{1}{L_S} \frac{\sum_{k}\sum_{n} c_k c^{*}_n e^{-it(\epsilon_k -\epsilon_n)} e^{i (p_k - p_n)(y-x)}}{\sum_{m} |c_m|^2} =  \frac{1}{L_S}\left| \sum_{k=0}^{d_S -1} c_k e^{-i\epsilon_k t}e^{ip_k(y-x)} \right|^2
\end{split}
\end{equation}
where, in the last step, we have considered again $\sum_{m=0}^{d_S-1}|c_m|^2 = 1$. Also in this case, from equation (\ref{76}) and considering $\ket{\psi(t,x)}_S= \sum_{k=0}^{d_S -1}c_k e^{-i\epsilon_k} e^{-ip_k}\ket{p_k}$, we can see that $P(y \: on\: S\:|\:x\:on\:R,\: t \: on \:C) =  \frac{1}{L_S}\left| \braket{y|\psi(t,x)} \right|^2$. 
\end{widetext}


\end{document}